\documentstyle[preprint,eqsecnum,aps]{revtex}

\begin{document}
\draft

\newcommand{\beq}{\begin{equation}}
\newcommand{\eeq}{\end{equation}}
\newcommand{\ba}{\begin{eqnarray}}
\newcommand{\ea}{\end{eqnarray}}
\newcommand{\boldsigma}{\mbox{\boldmath$\sigma$}}
\newcommand{\boldtau}{\mbox{\boldmath$\tau$}}
\newcommand{\boldomega}{\mbox{\boldmath$\omega$}}
\newcommand{\boldtheta}{\mbox{\boldmath$\theta$}}
\newcommand{\boldnabla}{\mbox{\boldmath$\nabla$}}
\newcommand{\comm}[1]{\left[#1\right]}

{\tighten

\title{Quantum Monte Carlo Studies of Relativistic Effects in Light Nuclei}
\author{J. L. Forest\cite{jlf}, V. R. Pandharipande\cite{vrp}}
\address{Department of Physics, University of Illinois at Urbana-Champaign,
1110 W. Green Street, Urbana, IL 61801}
\author{A. Arriaga\cite{aa}}
\address{Centro de Fisica Nuclear da Universidade de Lisboa,
	 Avenida Gama Pinto 2, 1699 Lisboa, Portugal}

\date{\today}

\maketitle

\begin{abstract}  
Relativistic Hamiltonians are defined as the sum of relativistic one-body
kinetic energy, two- and three-body potentials and their boost
corrections. In this work we use the variational Monte Carlo method
to study two kinds of relativistic effects in the binding energy
of $^3$H and $^4$He. The first is due to the nonlocalities in the
relativistic kinetic energy and relativistic one-pion exchange
potential (OPEP), and the second is from boost interaction. The OPEP
contribution is reduced by $\sim$ 15\% by the relativistic nonlocality, which
may also have significant effects on pion exchange currents.
However, almost all of this reduction is canceled by changes in the
kinetic energy and other interaction terms, and the total effect of
the nonlocalities on the binding energy is very small. The boost interactions,
on the other hand, give repulsive contributions of $\sim$ 0.4 (1.9) MeV
in $^3$H ($^4$He) and account for $\sim$ 37\% of the phenomenological part
of the three-nucleon interaction needed in the nonrelativistic Hamiltonians.
\end{abstract}


}

\section{Introduction}

It is generally accepted that QCD is the fundamental theory of strong
interactions, however, due to quark confinement, the genuine QCD
degrees of freedom are not explicit at low energies.
In low energy nuclear physics, nucleons and mesons are believed to be
the physical (effective) degrees of freedom. In the nonrelativistic
many-body theory, nuclei are regarded as bound states of nucleons
interacting via two- and three-body potentials.
All the sub-nucleonic and meson degrees of freedom, as well as relativistic
effects are, in some way, absorbed in these potentials. Typically the
nonrelativistic Hamiltonian is expressed as
\beq
H_{NR}=\sum_i\frac{p_i^2}{2m_i}+\sum_{i<j}v_{ij}+\sum_{i<j<k}V_{ijk}+\cdots,
\label{Hnr}
\eeq
and models of two- and three-body potentials are constructed by fitting 
observed data.  The
ellipsis in Eq.~(\ref{Hnr}) represents $N$-body interactions ($N \ge 4$) which
are thought to be much smaller than two- or three-nucleon interactions, and
therefore neglected.

The central problem is to solve the many-body Schr\"{o}dinger equation:
\beq
H_{NR}\:|\Psi\rangle=\:E|\Psi\rangle.
\label{sch-eq}
\eeq
The eigenvalues $E$ can be compared with experimental energies, and
the eigenstates $|\Psi\rangle$ can be used both to study the nuclear
structure and probe it through electron-nucleus scattering experiments, and to 
calculate rates of nuclear reactions which may have important applications
in several domains of physics.

Schr\"{o}dinger equation (\ref{sch-eq}) is difficult to solve due to
the large spin and isospin dependence of $v_{ij}$ and $V_{ijk}$.
Several techniques have been developed, among which are
Faddeev-Yakubovsky~\cite{Glockle93},
Harmonic-Hyperspherical basis~\cite{Kievsky93}, and Quantum Monte
Carlo (QMC)~\cite{Carlson91,Wiringa91} methods. The first two methods
are limited to solving 3- and 4-nucleon systems, whereas with the third 
method it is now possible to calculate the ground state energy and wave
function for $A$=2--8 nuclei with great accuracy.

Some of the results obtained by Pudliner {\it et al}.~\cite{Pudliner95}
are listed in Table I.  The Argonne $v_{18}$ 2-body
potential~\cite{Wiringa95}, fitted to $NN$ scattering data and the deuteron
binding energy, and Urbana IX 3-body potential~\cite{Pudliner95}
constrained to give the correct binding energy of $^3$H and density 
of nuclear matter, are used in these
calculations. It works rather well for $^4$He, however,
as can be seen from Table I, the  $A$=6, 7 and 8 nuclei appear to be
systematically underbound. It is interesting to note that a large fraction
of the total $v_{ij}$ comes from the one-pion exchange potential (OPEP)
and the dominant part of $V_{ijk}$ comes from two-pion exchange. Also notice
that the three-body interaction is much smaller than the two-body interaction,
yet it is crucial to obtain the observed energies, because of the large
cancelation between the kinetic energy and the two-body potential energy.

Although the nonrelativistic QMC techniques have advanced to such a level 
that the binding energies of light nuclei predicted by a realistic 
Hamiltonian can be calculated with $<$1\% error~\cite{Carlson91,Pudliner95},
the effective
description of nuclear dynamics by means of nonrelativistic Hamiltonians 
may have intrinsic deficiencies. In particular, when 
the nonrelativistic potentials are fit to the experimental data, relativistic
effects are automatically buried in these potentials. How well can these
effects be represented by means of local nonrelativistic potentials is
an important 
question to be answered. In other words, we may investigate whether an
explicit and more correct treatment of relativistic effects can resolve the
systematic underbinding of the nonrelativistic results for $A=6,7,8$ nuclei. 

Furthermore, with the recently completed multi GeV electron accelerator
facilities such as TJNAF, experiments will be performed at energy
and momentum transfer regimes where relativistic effects are substantial.
Clearly, investigation of these effects has become increasingly important.
Above all, no matter how small the relativistic effects might be,
understanding them is a fundamental quest, just like understanding the
fine and hyperfine structures in the hydrogen atom.

Several approaches have been developed to study various aspects of the
relativistic effects in few-body nuclear physics. They can be classified in two
main categories: effective field theories and relativistic Hamiltonian dynamics.
Within the first one, the Bethe-Salpeter equations for the  two- and three-body
systems have been solved using a separable kernel~\cite{Rupp92}. Also
covariant three-dimensional reductions of the relativistic integral equations
have been applied, along with one-boson exchange models for the kernels, to
the three-nucleon system. Here we refer to minimal relativity in the
Blankenbecler-Sugar equations~\cite{Sammarruca96} and, more recently, to 
the spectator (Gross) equations~\cite{Stadler97}. In the relativistic
Hamiltonian dynamics approach, relativistic covariance is achieved through
the Poincar\'e group theory. One application of this method is the light front
dynamics, which has been applied to the two-body
system~\cite{Keister91}, and the other is the instant form.

The present interest in the relativistic Hamiltonian dynamics in the instant  
form stems from the fact that the ground states of the Hamiltonian can be 
studied with the Quantum Monte Carlo methods that have already been developed in
the nonrelativistic approach. Earlier~\cite{Carlson93,Forest95II} and the
present study are limited to the $A$=2, 3 and 4 nuclei, but attempts to study
larger nuclei are in progress. In addition, the use of short-range
phenomenological terms in the interaction gives the flexibility to
allow a very good fit to the two-body scattering data with $\chi^2\sim 1$.
And finally, it is not obvious that the entire short and intermediate 
range 2-nucleon interaction can be represented as due to the exchange of 
a few types of mesons.  Thus more general ways of studying relativistic
effects in nuclei are desirable.

In this paper we report new results for the binding energies of the $A=3,4$
systems, 
using the relativistic Hamiltonian dynamics in the instant form, where for the
first time the nonlocalities induced by the relativistic effects in the
one-pion-exchange potential (OPEP) are taken into account.
In Sec. II we discuss the relativistic Hamiltonian used in this work.
In Sec. III we apply Variational Monte Carlo (VMC) techniques and present
results.  Finally we summarize in Sec. IV. Some of the detailed
derivations involved in this work are given in the Appendix.

\section{The Relativistic Hamiltonian}

In relativistic Hamiltonian dynamics in instant form the momentum (${\bf P}$)
and angular momentum (${\bf J}$) generators are chosen in the conventional
way and therefore are independent of interaction, while the
Hamiltonian ($H$) and boost (${\bf K}$) generators have interaction terms.
Based on the pioneering work of Bakamjian and Thomas~\cite{Bakamjian52} and
Foldy~\cite{Foldy61}, the relativistic Hamiltonian can be expressed as:
\begin{eqnarray}
H_{R}=\sum_i\left(\sqrt{m_i^2+p_i^2}-m_i\right)+\sum_{i<j}\left[
\tilde{v}_{ij}+{\delta}v_{ij}({\bf P}_{ij})\right]+
\sum_{i<j<k}\left[\tilde{V}_{ijk}+{\delta}V_{ijk}({\bf P}_{ijk})\right]\ 
+\cdots,
\label{Hr}
\end{eqnarray}
where $\tilde v_{ij}$ are two-body potentials in the ``rest frame'' of
particles $i$ and $j$ (i.e. the frame in which ${\bf P}_{ij}$=
${\bf p}_i$+${\bf p}_j$=0).
Similarly $\tilde V_{ijk}$ is the three-body potential in the frame in which
${\bf P}_{ijk}$=${\bf p}_i$+${\bf p}_j$+${\bf p}_k$=0.
The $\delta v_{ij}({\bf P}_{ij})$ and $\delta V_{ijk}({\bf P}_{ijk})$ are
called ``boost interactions'' and depend upon the total momentum of the
interacting particles. Obviously, $\delta v_{ij}({\bf P}_{ij}$=$0)$ and
$\delta V_{ijk}({\bf P}_{ijk}$=$0)$ vanish.
The $H_{NR}$ contains approximations
to the kinetic energy $T$, $\tilde{v}_{ij}$ and $\tilde{V}_{ijk}$,
and totally neglects the boost interactions. In the case of deuteron,
we can always go to its center of mass (c.m.) frame where total momentum
${\bf P}_{ij}$=0. The $\tilde{v}_{ij}$ is adequate to describe the deuteron
in its ``rest frame'', however, in $A>2$ nuclei the total momentum of
any pair of nucleons is not necessarily zero in the c.m. of the whole nucleus,
therefore the interaction between the pair can not be correctly described
by $\tilde{v}_{ij}$ alone.

Two kinds of relativistic effects in the interaction are studied in this
work: the boost interaction $\delta v_{ij}({\bf P}_{ij})$ due to the
motion of the c.m. of nucleons $i$ and $j$ in the rest frame of the whole
nucleus, and the nonlocality due to the relative motion of two nucleons
in their own c.m. frame. The latter will affect the ``rest frame'' potential
$\tilde{v}_{ij}$. The boost interaction has been studied in detail in
Refs. \cite{Forest95,Carlson93,Forest95II}, so we will only give a brief
discussion. On the other hand, the treatment of the nonlocality of OPEP
in quantum Monte Carlo calculations is discussed in detail.

The boost interaction $\delta v({\bf P}_{ij})$ is determined from the
``rest frame'' potential $\tilde{v}_{ij}$ through relativistic
covariance~\cite{Krajcik74,Friar75}. The $\delta v({\bf P}_{ij})$
is expanded in powers of $P_{ij}^2/4m^2$ and only the leading corrections
are considered in this work. The $\delta v({\bf P}_{ij})$ is given by:
\begin{equation}
\delta{v}({\bf P})=-\frac{P^2}{8m^2}\tilde{v}+\frac{i}{8m^2}\,\left[\ {\bf P}
\cdot{\bf r}{\bf P}\cdot{\bf p}, \tilde{v}\ \right]+
\frac{i}{8m^2}\,\left[\ (\mbox{\boldmath$\sigma$}_1-\mbox{\boldmath$\sigma$}_2)
\times{\bf P}\cdot{\bf p}, \tilde{v}\ \right],
\label{boost}
\end{equation}
where the subscripts $ij$ of $\tilde v$, ${\bf P}$, ${\bf p}$ and ${\bf r}$
have been suppressed for brevity. Here ${\bf p}$=$({\bf p}_i-{\bf p}_j)/2$
is the relative momentum operator, and $\boldsigma$=2${\bf s}$ are the
Pauli matrices for spin $1/2$ particles.

Various aspects of $\delta v({\bf P})$ are discussed in Ref.~\cite{Forest95}.
The first two terms of Eq.~(\ref{boost}) are denoted as $\delta v_{RE}$
and $\delta v_{LC}$; they have simple classical origins in
the relativistic energy-momentum relation and Lorentz contraction.
The last term contains contributions from Thomas precession
and quantum effects. They are denoted as $\delta v_{TP}$
and $\delta v_{QM}$ and are much smaller than the first two terms. 
For example, the contributions of $\delta v_{RE}$, $\delta v_{LC}$,
$\delta v_{TP}$ and $\delta v_{QM}$ to the energy of triton are
found~\cite{Carlson93} to be 0.23(2), 0.10(1), 0.016(2) and -0.004(2) MeV,
respectively. Since the main contribution comes from the first two terms,
for simplicity, we neglected the last two terms in the 3-, 4-body calculations
in this work.

In addition to the boost interaction, another source of relativistic
effects comes from the nonlocality.  In the following discussion we will
use the two-nucleon c.m. frame in which boost interaction vanishes,
and focus on the two-body ``rest frame'' interaction.
In most existing nonrelativistic potential models, the OPEP has been
calculated using the nonrelativistic Pauli spinors. 
Without $\pi NN$ form factors, it is given in momentum space by:
\beq
\tilde{v}_{\pi,NR}({\bf q})=-\frac{f_{\pi NN}^2}{\mu^2}\ \frac{\boldsigma_i
\cdot{\bf q}\boldsigma_j\cdot{\bf q}\boldtau_i\cdot\boldtau_j}
{\mu^2 +q^2},
\label{opepnr}
\eeq
where $f_{\pi NN}$ is the pion--nucleon coupling constant, $\mu$ is the
pion mass and ${\bf q}$ is the momentum transfer,
\beq
{\bf q} = {\bf p}-{\bf p}^{\prime}.
\eeq
Here ${\bf p}$ and ${\bf p}^{\prime}$ are the initial and final momenta of
nucleon $i$ in the center of mass frame, and the $\tilde{v}_{\pi,NR}$ is
local, i.e. it depends only on ${\bf q}$.

In contrast if one uses relativistic Dirac spinors the on-shell 
OPEP has the form:
\beq
\tilde{v}_{\pi,Rel}({\bf p}^{\prime},{\bf p})=\frac{m}{\sqrt{m^2+p^{\prime 2}}}
\ \tilde{v}_{\pi,NR}({\bf q})\ \frac{m}{\sqrt{m^2+p^2}}.
\label{opep}
\eeq
This potential is dependent not only on ${\bf q}$ but also on ${\bf p}$
and ${\bf p}^{\prime}$,
which results in a nonlocal potential in the configuration space.
The interaction (\ref{opep}) is regarded as energy independent and used
in many-body Schr\"{o}dinger equations. By expanding the square roots it can
be easily verified that the leading correction
$(\tilde{v}_{\pi,Rel}-\tilde{v}_{\pi,NR})$ is of order $p^2/m^2$, i.e.,
of order $v^2/c^2$ where $v$ denotes the velocity of the nucleons in the
center of mass frame.

In Ref. \cite{Forest95} it is shown that the relation between the boost
interaction 
$\delta v({\bf P})$ and the static $\tilde{v}_{NR}$ is independent of the 
origin of $\tilde{v}_{NR}$ up to order $P^2/m^2$, and presumably beyond. 
Thus the knowledge of the static $\tilde{v}_{NR}$ is sufficient to 
obtain $\delta v({\bf P})$.  In contrast the above relation between 
$\tilde{v}_{\pi ,Rel}$ and $\tilde{v}_{\pi ,NR}$ (Eq. \ref{opep})
is specific for the 
interaction due to exchange of pseudoscalar mesons by Dirac nucleons 
via either pseudoscalar or pseudovector coupling.  By expanding the 
square roots in Eq. (\ref{opep}) we obtain:
\beq
\tilde{v}_{\pi ,Rel}({\bf p}^{\prime},{\bf p}) = 
\tilde{v}_{\pi ,NR}({\bf q}) \left(1-\frac{p^{\prime 2}}{2 m^2} 
-\frac{p^2}{2 m^2} + \cdots \right),
\eeq
whereas the interactions generated by exchange of scalar $(S)$ or 
vector $(V)$ mesons have different relations~\cite{Forest95}:
\ba
\tilde{v}_{S ,Rel}({\bf p}^{\prime},{\bf p}) &=& 
\tilde{v}_{S ,NR}({\bf q}) \left(1
-\frac{({\bf p}^{\prime}+{\bf p})^2}{2 m^2} + \cdots \right), \\
\tilde{v}_{V ,Rel}({\bf p}^{\prime},{\bf p}) &=& 
\tilde{v}_{V ,NR}({\bf q}) \left(1
+\frac{({\bf p}^{\prime}+{\bf p})^2}{2 m^2} + \cdots \right).
\ea

Realistic models of nuclear forces contain 
momentum dependent terms which presumably take into account 
some of the relativistic corrections to the phenomenological 
short and intermediate range parts of $\tilde{v}_{NN}$.  However, 
most configuration space models do not contain long range, momentum dependent 
tensor forces occuring in $\tilde{v}_{\pi ,Rel}$.  In exact calculations 
the tensor force can not be generally expanded in powers of $p^2/m^2$. 
In any truncated expansion the force diverges at large values of $p$, 
and can yield divergent attraction.  

In the present work the two-nucleon potential is expressed as:
\beq
\tilde{v}_{NN}=\tilde{v}_{\pi,Rel}+\tilde{v}_R
\label{vnn}
\eeq
where $\tilde{v}_R$ is the remaining part of the 2-body potential which is
phenomenological. We can also
write the OPEP given in Eq.\ (\ref{opep}) as:
\beq
\tilde{v}_{\pi,Rel}=\tilde{v}_{\pi,NR}\ +\ \left(\tilde{v}_{\pi,Rel}-
\tilde{v}_{\pi,NR}\right).
\eeq
The term in parenthesis is the relativistic correction. The nonrelativistic
potential models do not consider this correction explicitly: the data is fit
using $\tilde{v}_{\pi,NR}$ in Eq.\ (\ref{vnn}), thus some of its effects
go into the phenomenological part of the potential $\tilde{v}_R$. The
$\tilde{v}_R$ in relativistic Hamiltonian differs from that in nonrelativistic
Hamiltonian due to the difference in $\tilde{v}_{\pi,Rel}$ and
$\tilde{v}_{\pi,NR}$ as well as that in the kinetic energy operators.

We construct our $H_R$ to be phase equivalent to the
isoscalar part of the nonrelativistic $H_{NR}$ containing
Argonne $v_{18}$. The relativistic effects can then be studied by comparing
results obtained from our $H_R$ and the isoscalar $H_{NR}$ without
considering the small isospin symmetry breaking terms in the latter.
The relativistic Hamiltonian for two-nucleon system in its center of mass
frame is chosen as:
\beq
H_R=2\sqrt{p^2+m^2}-2m\ +\ \frac{m}{\sqrt{m^2+p^{\prime 2}}}\
\tilde{v}_{\pi,NR}({\bf q})\ \frac{m}{\sqrt{m^2+p^2}}\ +\ \tilde{v}_R.
\eeq
where $\tilde{v}_R$ has the same form as the isoscalar part of Argonne
$v_{18}$~\cite{Wiringa95}:
\beq
\tilde{v}_R=\sum_{p=1,14}v_p(r_{ij})O_{ij}^p,
\label{av18op}
\eeq
\ba
O_{ij}^{p=1,14}&=&1, \boldtau_i\cdot\boldtau_j, \boldsigma_i\cdot\boldsigma_j,
(\boldsigma_i\cdot\boldsigma_j)(\boldtau_i\cdot\boldtau_j), S_{ij},
S_{ij}(\boldtau_i\cdot\boldtau_j), {\bf L}\cdot{\bf S},
{\bf L}\cdot{\bf S}(\boldtau_i\cdot\boldtau_j), \nonumber \\
& & L^2, L^2(\boldtau_i\cdot\boldtau_j), L^2(\boldsigma_i\cdot\boldsigma_j),
L^2(\boldsigma_i\cdot\boldsigma_j)(\boldtau_i\cdot\boldtau_j),
({\bf L}\cdot{\bf S})^2, ({\bf L}\cdot{\bf S})^2(\boldtau_i\cdot\boldtau_j).
\ea
The $\tilde{v}_{\pi,NR}$ used in Argonne $v_{18}$ is given by:
\beq
\tilde{v}_{\pi,NR}({\bf r}) = \frac{1}{3}\mu \frac{f_{\pi NN}^2}{4\pi}
\left[Y_{\pi}(r)\boldsigma_i\cdot\boldsigma_j+T_{\pi}(r)S_{ij}\right]
\boldtau_i\cdot\boldtau_j,
\label{av18}
\eeq
where
\ba
Y_{\pi}(r)&=&\frac{e^{-\mu r}}{\mu r}\left(1-e^{-cr^2}\right), \\
T_{\pi}(r)&=&\left(1+\frac{3}{\mu r}+\frac{3}{(\mu r)^2}\right)
\left(1-e^{-cr^2}\right)^2, \\
S_{ij}&=&3\boldsigma_i\cdot\hat{r}\boldsigma_j\cdot\hat{r}-\boldsigma_i\cdot
\boldsigma_j.
\label{more-av18}
\ea
We note that this $\tilde{v}_{\pi,NR}$ does not contain the part of
$\boldsigma_i\cdot\boldsigma_j\boldtau_i\cdot\boldtau_j$ interaction
which acquires a $\delta(r_{ij})$ function form in the limit of point
particles. This $\delta-$function is probably spread out by the finite
size of the nucleons, and contributes to the short range part of
$v_{NN}$. However, it is difficult to extract it from the
phenomenological models. Moreover the dominant
contribution and the nonlocality effect seem to come from the tensor
part of OPEP.

Recently, after completion of the present work, Kamada and Gl\"{o}ckle 
\cite{KG98} found an elegant method to obtain a potential $v_{Rel}$ that 
gives exactly the same phase shifts with relativistic kinetic 
energy that a known $v_{NR}$ gives with nonrelativistic kinetic 
energy.  Our objective here is not just to find a $v_{Rel}$ that is 
phase equivalent to the Argonne $v_{18}$; we additionally 
require it to have the 
$\tilde{v}_{\pi,Rel}$ long range behavior.  Both relativistic and
nonrelativistic models of $\tilde{v}$ contain theoretical long range
OPEP; the scattering data is used to determine only the phenomenological 
part $\tilde{v}_R$ in these interactions.  Our two models are not as
exactly phase equivalent as Kamada and Gl\"{o}ckle's $v_{Rel}$ and $v_{NR}$ 
are, however the differences in their phase shifts are negligibly small 
compared to the uncertainties in the Nijmegen phase shifts \cite{SKRS93}.

The parameters of the function $v_p(r_{ij})$ of $\tilde{v}_R$ are
obtained by fitting
the phase shifts and deuteron binding energy.
Traditionally phase shifts are calculated in configuration space,
however, in the relativistic case, the Hamiltonian
contains $\sqrt{m^2-\nabla^2}$ which is nonlocal in configuration space,
therefore we calculate them in momentum space. The details of the
momentum-space technique have been discussed in Ref.~\cite{Glockle86}.

Some of the important phase shifts are plotted in Fig.~\ref{fig:phase}. 
The diamond symbols represent the reference nonrelativistic phase shifts
obtained with $H_{NR}$, the plus symbols represent those calculated from
$H_R$ before re-adjusting the parameters in $v_p(r)$, and the square
symbols correspond to those after. The reference phase shifts are almost
exactly reproduced by the relativistic Hamiltonian $H_R$ as indicated by
the good overlap of the diamond and square symbols in Fig.~\ref{fig:phase}.
The deviations between the plus
and diamond symbols reflect the total effect of replacing nonrelativistic
kinetic energy and $\tilde{v}_{\pi,NR}$ by the relativistic kinetic energy and
$\tilde{v}_{\pi,Rel}$ in $H_{NR}$, and are not too large
except for the mixing parameter $E_1$ of $^3S_1-^3\!D_1$. This indicates that
relativistic nonlocal effects in two nucleon scattering at $E_{\mbox{lab}}<400$
MeV are rather small. $E_1$ is primarily determined by the tensor force;
the relatively large change in $E_1$ is due to the nonlocality
of the tensor force in $\tilde{v}_{\pi,Rel}$.

The new two-body potential is essentially phase equivalent
to isoscalar part of Argonne $v_{18}$ and predicts similar deuteron properties
as listed in Table II. Note that the present $H_{NR}$ and $H_R$ do not
contain electromagnetic interactions. The experimental value of deuteron
binding energy (-2.224 MeV) can be obtained from the full Argonne $v_{18}$ with
electromagnetic interactions.
The 14 operator components of the relativistic and nonrelativistic potentials
are compared in Fig.~\ref{fig:v14}.
Only the static part of the relativistic OPEP, obtained by setting 
$m/\sqrt{p^2+m^2}$ equal to unity is used in Fig.~\ref{fig:v14}.
Since it is the same as the nonrelativistic OPEP, the difference between the
potentials shown in the figure is entirely due to that in the phenomenological 
part $\tilde{v}_R$.  Equation (\ref{opep}) shows that
the $\tilde{v}_{\pi,Rel}$ is smaller than the $\tilde{v}_{\pi,NR}$ for 
$p$ or $p^{\prime} \neq 0$.

The deuteron $S$- and $D$-wave functions are shown in Fig.~\ref{fig:wavefunc}.
Here the relativistic $D$-wave is slightly smaller than the nonrelativistic one,
presumably because the relativistic tensor potential
is smaller than the nonrelativistic tensor potential (Eq.~\ref{opep}) for
large $p$ and $p^{\prime}$.
The deuteron wave functions in momentum space are shown in
Fig.~\ref{fig:momdist}.  Note that the relativistic wave functions are
not very different from the nonrelativistic ones.  The ratio of the 
two momentum space D-wave functions can be easily understood as 
discussed below.

The exact ground state wave function $\Psi$ can be expanded in a 
complete set of states $|i\rangle$:
\beq
|\Psi\rangle = \sum_i \phi_i |i \rangle.
\eeq
For a Hamiltonian given by $T+v$, where $|i\rangle$ are eigenstates of 
$T$, the amplitudes $\phi_i$ are given by
\beq
\phi_i = - \frac {\langle i|v|\Psi \rangle} 
{\langle i|T-E_0| i \rangle},
\label{alpi}
\eeq
as can be verified from the Schr\"{o}dinger equation
$H |\Psi \rangle = E_0 |\Psi \rangle$.  Here $E_0$ is the ground state energy,

In the case of the deuteron 
we can choose $T$ as the kinetic energy operator and $|i\rangle$ as 
S- and D-waves with momentum $p$, denoted by $|p_l\rangle$, for $l=S,D$.
The amplitudes $\phi_l(p)$ of these waves give the deuteron wave 
function in momentum space.  We can estimate the difference between 
the nonrelativistic and relativistic deuteron D-state wave function 
at large momentum by assuming that it is primarily generated by the 
OPEP.  In the nonrelativistic case this gives:
\beq
\phi_{D,NR}(p) = - \frac{m}{p^2}\langle p_D |\tilde{v}_{\pi,NR}| \Psi \rangle,
\eeq
where we have neglected the $E_0$ in the denominator of Eq.~(\ref{alpi}), 
since it is much smaller than the kinetic energy $p^2/m$ at large $p$.  
In the relativistic case
\beq
\phi_{D,Rel}(p) = - \frac{1}{2(\sqrt{m^2+p^2}-m)} \frac{m}{\sqrt{m^2+p^2}}
\langle p_D |\tilde{v}_{\pi,NR} | \Psi \rangle,
\eeq
where the first factor is the relativistic 
kinetic energy denominator, and the second comes from the 
$m/E^{\prime}$ factor in the $\tilde{v}_{\pi,Rel}$ (Eq.~\ref{opep}). 
The other $m/E$ factor in the $\tilde{v}_{\pi,Rel}$ operates on the 
$\Psi$.  It is set to unity because most of the deuteron wave function 
has small relative momenta.

Neglecting the small difference between 
the relativistic and nonrelativistic $\Psi$, 
the ratio of the $\phi_D(p)$ is found to be
\beq
\frac{\phi_{D,Rel}(p)}{\phi_{D,NR}(p)} = \frac{p^2}
{2(\sqrt{m^2+p^2}-m) \sqrt{m^2+p^2}}.
\label{ratio}
\eeq
The above estimate is fairly close to the ratio of the 
calculated D-wave functions as can be seen in Fig.~\ref{fig:ratio}.
Note that this ratio is 
smaller if the $\tilde{v}_{\pi,Rel}$ is used with the nonrelativistic 
kinetic energy in $\phi_{D,Rel}(p)$ (dotted line), and it is larger than
one when the $\tilde{v}_{\pi,NR}$ is used with the relativistic kinetic
energy (dot-dashed line).  The relativistic corrections 
to the interaction and kinetic energies have opposite effects on 
the wave function.  The
difference between the S-wave functions is influenced by the changes in 
the kinetic energy and the $\tilde{v}_R$.  The effects of these changes 
on the phase shifts and the deuteron energy must cancel by construction, 
and they seem to largely cancel in the $\phi_S(p)$.

Some of the deuteron momentum space results are listed in Table III.
This table offers a microscopic view of how various relativistic effects
were buried in the nonrelativistic models.
Relativistic nonlocalities reduce the OPEP contribution by $\sim$2.6 MeV,
while the relativistic kinetic energy is smaller by $\sim -1$ MeV,
giving a net effect of
1.6 MeV which is canceled by the change in the phenomenological $\tilde{v}_R$.

The variational Monte Carlo calculations for the
$A>2$ systems have to be carried out in configuration space. We therefore
have to Fourier transform the $\tilde{v}_{\pi,Rel}$ which depends on
both ${\bf p}$ and ${\bf p}^{\prime}$ (Eq.~\ref{opep}), yielding a
nonlocal potential in configuration space:
\beq
\tilde{v}_{\pi,Rel}({\bf r}^{\prime},{\bf r}) =
\int \frac{\mbox{d}^3p}{(2\pi)^3}
\frac{\mbox{d}^3p^{\prime}}{(2\pi)^3}\ \mbox{e}^{-i{\bf p}^{\prime}\cdot
{\bf r}^{\prime}}\ \tilde{v}_{\pi,Rel}({\bf p}^{\prime},{\bf p})\ 
\mbox{e}^{i{\bf p}\cdot{\bf r}}.
\label{fourier}
\eeq
The exact integral in Eq.\ (\ref{fourier}) is extremely difficult to calculate.
The series obtained by expanding
$\tilde{v}_{\pi,Rel}({\bf p}^{\prime},{\bf p})$ in powers of $p^2/m^2$
is given by:
\beq
\tilde{v}_{\pi,Rel}({\bf p}^{\prime},{\bf p}) = \tilde{v}_{\pi,NR}({\bf q})
\left(1-\frac{p^2+{p^{\prime}}^2}{2m^2}
+\frac{3p^4+2p^2{p^{\prime}}^2+3{p^{\prime}}^4}{8m^4}+\cdots\right)
\eeq
However, this series does not have good convergence. In the case of deuteron,
the expectation value of $\tilde{v}_{\pi,NR}$, for the eigenfunction of 
our relativistic Hamiltonian is -21.39 MeV.
The term in $\tilde{v}_{\pi,Rel}$, of order $1/m^2$,
contributes 3.48 MeV to the
$\langle\tilde{v}_{\pi,Rel}\rangle$,
while that of order $1/m^4$ gives -1.45 MeV, and the exact 
$\langle\tilde{v}_{\pi,Rel}-\tilde{v}_{\pi,NR}\rangle$ is 2.59 MeV.
Therefore the series converges slowly to the exact value.
This may appear surprising because the expectation value of the kinetic
energy of the deuteron (Table III) is only about 20 MeV, giving
$p^2/m^2\approx 0.02$ on average.  However, the deuteron has large 
momentum components via the D-wave, or equivalently the tensor 
correlations, and most of the OPEP contribution is from those. 
Thus it is not surprising that the expectation value of OPEP is 
sensitive to higher powers of nucleon velocities.

For the relativistic OPEP, a good convergence is achieved by using the
variables:
\ba
{\bf Q} &=& \frac{1}{2}({\bf p}+{\bf p}^{\prime}), \ \ \ \ \ \ {\bf q} =
{\bf p}-{\bf p}^{\prime}, \label{var1} \\
{\bf x} &=& \frac{1}{2}({\bf r}+{\bf r}^{\prime}), \ \ \ \ \ \ \ {\bf y} =
{\bf r}-{\bf r}^{\prime},
\label{var2}
\ea
for which
\ba
\tilde{v}_{\pi,Rel}&=&\tilde{v}_{\pi,NR}({\bf q})\frac{m^2}{\sqrt{
\left(m^2+Q^2+q^2/4\right)^2-({\bf Q}\cdot{\bf q})^2}}, \nonumber \\
&=&\tilde{v}_{\pi,NR}({\bf q})\left[1-\frac{Q^2+\frac{q^2}{4}}{m^2+Q^2+
\frac{q^2}{4}}+\frac{1}{2}\frac{m^2({\bf Q}\cdot{\bf q})^2}
{\left(m^2+Q^2+\frac{q^2}{4}\right)^3}+\cdots\right].
\label{expansion}
\ea
Here we expanded the $\tilde{v}_{\pi,Rel}$ in powers of
$({\bf Q}\cdot{\bf q})^2/(m^2+Q^2+q^2/4)^2$.
This series appears to converge rapidly. In the case of deuteron,
the leading relativistic correction given by the second term is 2.7 MeV, 
the third term gives -0.18 MeV, and the exact value is 2.59 MeV.
The third term contains
$\theta_{Qq}$ dependence and results in complicated operator forms as shown
in the Appendix. Moreover, the third and higher terms account for only 
$\sim 4$ \% of the relativistic correction to OPEP expectation value in 
the deuteron ({\em i.e.} $\sim 0.6$ \% of $\langle \tilde{v}_{\pi,Rel} 
\rangle$).  Therefore only the first two terms are considered
in this work, and the relativistic OPEP is approximated by:
\beq
\tilde{v}_{\pi,Rel} = \frac{m^2}{m^2+Q^2+\frac{q^2}{4}}\ \tilde{v}_{\pi,NR}
({\bf q}).
\label{opepnew}
\eeq
With this $\tilde{v}_{\pi,Rel}$ in Eq.~(\ref{vnn}), we refit the phase
shifts and deuteron binding energy.  The parameters in $v_p(r)$ are very
similar to those obtained with the exact $\tilde{v}_{\pi,Rel}$ given by
Eq.~(\ref{opep}). The new phase shifts and the relativistic potentials
are very similar to those shown in Figs. \ref{fig:phase} and \ref{fig:v14}.

The configuration space potential (Eq.~\ref{fourier}) is given by:
\beq
\tilde{v}_{\pi,Rel}({\bf x},{\bf y})=\int\frac{\mbox{d}^3 Q}
{(2\pi)^3}\frac{\mbox{d}^3 q}{(2\pi)^3}\,\frac{m^2}
{m^2+Q^2+\frac{q^2}{4}}\,\tilde{v}_{\pi,NR}({\bf q})
\,\mbox{e}^{i({\bf Q}\cdot{\bf y}
+{\bf q}\cdot{\bf x})},
\label{integral}
\eeq
and is simple to evaluate. The $\tilde{v}_{\pi,NR}$
(Eq.~\ref{av18}) in momentum space is given by:
\ba
\tilde{v}_{\pi,NR}({\bf q})&=&\int \tilde{v}_{\pi,NR}({\bf r})
\,e^{i{\bf q}\cdot{\bf r}}\,\mbox{d}^3r, \nonumber \\
&=&\frac{1}{3}\mu f_{\pi NN}^2\left[
{\cal Y}_{\pi}(q)\,\boldsigma_i\cdot\boldsigma_j
+{\cal T}_{\pi}(q)\left(\boldsigma_i\cdot\boldsigma_j-
\frac{3}{q^2}\boldsigma_i\cdot{\bf q}\boldsigma_j\cdot{\bf q}\right)\right]
\boldtau_i\cdot\boldtau_j,
\label{vpift}
\ea
where
\ba
{\cal Y}_{\pi}(q)&=&\int Y_{\pi}(r)j_0(qr)r^2\mbox{d}r,
\label{yft} \\
{\cal T}_{\pi}(q)&=&\int T_{\pi}(r)j_2(qr)r^2\mbox{d}r.
\label{tft}
\ea
Substituting these into Eq.~(\ref{integral}) gives:
\beq
\tilde{v}_{\pi,Rel}(x,y)=\frac{1}{3}\mu \frac{f_{\pi NN}^2}{4\pi}\,f(y)\,\left[
F_{\sigma\tau}(x,y)
\ \boldsigma_i\cdot\boldsigma_j
+F_{t\tau}(x,y)\ S_{ij}(\hat{x},\hat{x})\right]\boldtau_i\cdot\boldtau_j,
\label{vnonlocal}
\eeq
with
\ba
F_{\sigma\tau}(x,y) &=& \frac{2}{\pi}\int q^2\mbox{d}q\ 
{\cal Y}_{\pi}(q)\ j_0(qx)\ \mbox{e}^{-(\sqrt{m^2+q^2/4}-m)\ y}, \\
F_{t\tau}(x,y) &=& \frac{2}{\pi}\int q^2\mbox{d}q\ {\cal T}_{\pi}(q)\ j_2(qx)\ 
\mbox{e}^{-(\sqrt{m^2+q^2/4}-m)\ y}, \\
f(y) &=& \frac{m^2}{4\pi}\frac{e^{-my}}{y}, \\
S_{ij}(\hat{x},\hat{x}) &=& 3\,\boldsigma_i\cdot\hat{{\bf x}}\,\boldsigma_j\cdot
\hat{{\bf x}}-\boldsigma_i\cdot\boldsigma_j.
\label{Sijx}
\ea
In the limit $m\rightarrow\infty$, $f(y)$ becomes $\delta^3({\bf y})$,
$F_{\sigma\tau}(x,y)\rightarrow Y_{\pi}(x)$ and
$F_{t\tau}(x,y)\rightarrow T_{\pi}(x)$.  When ${\bf y}\rightarrow 0$,
we have ${\bf r}^{\prime}={\bf r}$, ${\bf x}={\bf r}$
and Eq.~(\ref{vnonlocal}) becomes $\tilde{v}_{\pi,NR}$.
Figure~\ref{fig:f}
shows $F_{\sigma\tau}(x,y)$ and $F_{t\tau}(x,y)$ as a function of $x$ for
various values of $y$. Note that the solid lines for $y=0$
correspond to the nonrelativistic $Y_{\pi}(x)$ and $T_{\pi}(x)$.
The volume integral of $F_{\sigma\tau}(x,y)$ is independent of y, whereas
that of $F_{t\tau}(x,y)$ decreases with y. Therefore relativistic
effects mostly come from the tensor part of OPEP.

\section{Variational Monte Carlo calculations and results}

\subsection{VMC techniques}

With the relativistic Hamiltonian discussed in the previous section,
we can proceed to evaluate the energy expectation value
\beq
\langle H_R\rangle=\langle T\rangle+\langle \tilde{v}_{ij}\rangle
+\langle\delta v_{ij}\rangle +\langle \tilde{V}_{ijk}\rangle
\label{vmc}
\eeq
for $A\ge 3$ nuclei using the Monte Carlo technique.

The Monte Carlo method~\cite{Kalos86} offers a useful way to handle
the multidimensional integrals which would otherwise be impractical
by the usual numerical methods. The basis of this method is that
instead of integrating over a regular array of points, we 
sum over a set of configurations $\{ {\bf R}_i \}$ distributed with
probability $w({\bf R})$. 
Here ${\bf R}=({\bf r}_1,{\bf r}_2,\cdots,{\bf r}_A)$ denotes the 
configuration of all the nucleons in the nucleus.
There are various techniques for sampling
$w({\bf R})$~\cite{Kalos86}, and in this work Metropolis
sampling method~\cite{Metropolis53} is used to treat the complicated
distributions.

Variational Monte Carlo (VMC) technique is based on variational principle
that the minimum expectation value of the Hamiltonian is closest to
the ground state energy of the system. Starting from 
a variational wave function, which depends upon several
variational parameters $(\alpha_1,\alpha_2,\cdots,\alpha_n)$,
we evaluate the expectation value of the Hamiltonian
using the Monte Carlo configuration samples ${\bf R}_i$:
\ba
\langle\hat{H} \rangle &=& \frac{\int \mbox{d}{\bf R} \Psi_v^{\dagger}({\bf R})
\hat{H}\Psi_v({\bf R})}{\int\mbox{d}{\bf R}\Psi_v^{\dagger}({\bf R})\Psi_v
({\bf R})} \nonumber \\
&=&\frac{\frac{1}{N_c}\sum_{i=1}^{N_c}\left(\Psi_v^{\dagger}({\bf R}_i)
\hat{H} \Psi_v({\bf R}_i)\right)/w({\bf R}_i)}
{\frac{1}{N_c}\sum_{i=1}^{N_c}\left(\Psi_v^{\dagger}({\bf R}_i)
\Psi_v({\bf R}_i)\right)/w({\bf R}_i)} \pm \delta,
\ea
where $\delta$ is the standard deviation. Typically block averaging scheme
is used to obtain a normal distribution and the error can be conveniently
evaluated from it. We divide $N_c$ configurations into $N_b$ blocks each
containing $N_0=N_c/N_b$ configurations. The average
\beq
\tilde{H}_b = \frac{\frac{1}{N_0}\sum_{i=1}^{N_0}\left(\Psi_v^{\dagger}({\bf R}_
i)
\hat{H} \Psi_v({\bf R}_i)\right)/w({\bf R}_i)}
{\frac{1}{N_0}\sum_{i=1}^{N_0}\left(\Psi_v^{\dagger}({\bf R}_i)
\Psi_v({\bf R}_i)\right)/w({\bf R}_i)}
\eeq
is evaluated for each block. The expectation value of $H$ is given by
\beq
\langle H \rangle = \frac{1}{N_b}\sum_{i=1}^{N_b}\tilde{H}_b
\eeq
with the standard deviation
\beq
\delta = \frac{1}{N_b}\sqrt{\sum_{i=1}^{N_b}\left(\tilde{H}_b-\langle H\rangle
\right)^2}.
\eeq
The Monte Carlo result is exact when the number of configurations
$N_c\rightarrow\infty$,
although in practice $N_c$=50000 seems to be enough to obtain results
with sufficiently small statistical errors.
The weight function is usually chosen to be
\beq
w({\bf R}_i)=\Psi_v^{\dagger}({\bf R}_i)\Psi_v({\bf R}_i)
\eeq
to maintain small Monte Carlo error. Note that when $\Psi_v$ is the eigenstate
of $H$, the Monte Carlo sampling error becomes zero.
Finally, the parameters $(\alpha_1,\alpha_2,\cdots,\alpha_n)$
are varied to minimize the energy.

Some of the terms in $\langle H_R\rangle$ (Eq.\ \ref{vmc}) can be
calculated straightforwardly and
have been discussed in Refs.~\cite{Arriaga95} and \cite{Forest95II}.
The terms that require special techniques are the relativistic
kinetic energy $\langle\sqrt{m^2-\nabla^2}\rangle$ 
and $\tilde{v}_{\pi,Rel}$ in the two-body potential.
The kinetic energy term has been calculated previously in
Ref.~\cite{Carlson93}, and the calculation of
$\langle \tilde{v}_{\pi,Rel} \rangle$ is discussed in Sec.\ IIIC.

\subsection{Relativistic wave functions}
In the nonrelativistic case, variational wave functions of the form
\beq
|\Psi_v \rangle = \left(1+\sum_{i<j<k}F_{ijk}\right)\,\left({\cal S}
\prod_{i<j}F_{ij}\right)\,|\Phi\rangle,
\eeq
having symmetrized product of pair correlation operators $F_{ij}$ and
a sum of triplet correlations $F_{ijk}$
operating on an antisymmetric, uncorrelated wave function
$|\Phi\rangle$, have been commonly used. 
The $F_{ij}$ and $F_{ijk}$ correlation operators
reflect the effects of two-, three-body interactions
on the wave function.
The uncorrelated wave function has no spatial dependence
for $A\le$4 nuclei.
A good representation of such a wave function is given in 
Ref.~\cite{Arriaga95}.

The pair correlation operator $F_{ij}$ is constructed from correlation
functions which satisfy Schr\"{o}dinger-like 2-body equations, with appropriate
boundary conditions. Their solutions are like deuteron wave functions
$\Psi_{NR}$ and $\Psi_R$ displayed in Fig.~\ref{fig:wavefunc}. In the case
of $A$=2 deuteron, both nonrelativistic and relativistic correlation functions
can be easily solved in momentum space; they are not very different
from each other as can be seen in Fig.~\ref{fig:wavefunc}.
In $A>$2 nuclei, the nonrelativistic pair correlation 
equations can be easily solved in configurations space,
however, the relativistic equations are more difficult to solve.
Therefore we seek good
approximations for the relativistic pair correlation functions.

Our method can be easily illustrated using the example of deuteron. Its
variational wave function is expressed as:
\beq
\Psi=\Psi_{NR}+\lambda (\Psi_R-\Psi_{NR}),
\eeq
$\langle\Psi|H_R|\Psi\rangle$ is calculated using VMC, and 
$\lambda$ is varied 
to minimize it. The results are shown in Fig.~\ref{fig:deutvmc}.
The error bars shown in Fig.~\ref{fig:deutvmc} originate from the statistical
sampling and are $<$1\% of the binding energy. The same configurations are
used to calculate the energies for all $\lambda$, hence the errors are
correlated. The minimum value
of $\langle H_R\rangle$ does occur at $\lambda=1$ where $\Psi=\Psi_R$
as expected. The difference in $\langle H_R\rangle$ between $\lambda=0$
(using a nonrelativistic wave function) and $\lambda=1$ (using a
relativistic one) is $\sim$0.04 MeV. This means that if we were
to use the nonrelativistic wave function to calculate the expectation
value of $H_R$, the result will be off by only 2\% for the deuteron.

In heavier nuclei we also expect the optimum nonrelativistic wave function
to provide a good approximation for relativistic wave function. The
difference between the two is presumably largest in
$^3S_1-^3\!D_1$ and $^1S_0$ correlation functions at small $r$.
We therefore define:
\ba
f_{0,1}^c &=& f_{0,1,NR}^c\ (1\ +\ \lambda\ \xi_{\,^1\!S_0}), \label{f01c} \\
f_{1,0}^c &=& f_{1,0,NR}^c\ (1\ +\ \lambda\ \xi_{\,^3\!S_1}), \label{f10c} \\
f_{1,0}^t &=& f_{1,0,NR}^t\ (1\ +\ \lambda\ \xi_{\,^3\!D_1}), \label{f10t}
\ea
where $\xi_c$ for the channel $c$ is defined as:
\beq
\xi_c(r)=\frac{\phi_{c,R}(r)-\phi_{c,NR}(r)}{\phi_{c,NR}(r)}.
\eeq
At small $r$, the central $f^c_{1,0}$ and tensor $f^t_{1,0}$ correlation
functions in spin-isospin $S,T = 1,0$ states are proportional
to the S and D radial wave functions of the deuteron, respectively.
Therefore the $\xi_{^3S_1}$ and $\xi_{^3D_1}$ can be calculated exactly
in momentum space using the relativistic and nonrelativistic 
deuteron wave functions for the $\phi_{c,R}$ and $\phi_{c,NR}$.
These $\xi$'s are rather
short-ranged (Fig.~\ref{fig:diffall}), and
we do not expect them to vary significantly in larger nuclei.
Wiringa~\cite{Wiringa91} has shown that the nonrelativistic
correlation functions for $^2$H, $^3$H and $^4$He are almost the same 
at small $r$.

A similar calculation of $\xi_{^1S_0}$ is not possible
because of the absence of a bound state in that channel. However, we can
obtain an artificial $^1S_0$ bound state by slightly increasing the 
strength of intermediate range attraction in $v(^1S_0)$.
The binding energy and the wave function at large $r$ are very sensitive
to small changes in $v(^1S_0)$, however, $\xi_{^1S_0}$ is relatively
insensitive. As an example, the $\xi_{^1S_0}$ obtained from artificial
$^1S_0$ bound states with energies of -1 and -10 MeV, shown 
in Fig.~\ref{fig:diffall}, are very similar.

The VMC energy of $^3$H with the relativistic Hamiltonian is shown as a
function of $\lambda$ in Fig.~\ref{fig:tritonvmc}. The minimum occurs
at $\lambda$=0.5 instead of 1 (the expected value), however, the
difference in $\langle H_R\rangle$ between $\lambda$=0.5 and 1 is rather
small and of the order of the Monte Carlo sampling error.  The minimum
energy for $^4$He occurs at the expected $\lambda$=1.0.

\subsection{Expectation value of the nonlocal potential}

Consider an $A$-nucleon system whose wave function is denoted as
$\Psi({\bf r}_1,{\bf r}_2,\cdots,{\bf r}_A)$.
We define:
\ba
{\bf x}_i = \frac{1}{2}({\bf r}_i+{\bf r}_i^{\prime}),\ \ \ \ \ \ 
{\bf x}_j = \frac{1}{2}({\bf r}_j+{\bf r}^{\prime}_j),
\ea
so that
\ba
{\bf x} = {\bf x}_i-{\bf x}_j, \ \ \ \ \ \ 
\frac{{\bf y}}{2} = {\bf r}_i-{\bf r}_i^{\prime}={\bf r}_j^{\prime}-{\bf r}_j,
\ea
as illustrated in Fig.~\ref{fig:nonlocal}. The expectation value of
relativistic OPEP is then given by
\ba
\langle \tilde{v}_{\pi,Rel} \rangle &=& \sum_{i<j}\int \prod_{k\ne i,j}
\mbox{d}^3 r_k\mbox{d}^3x_i\mbox{d}^3x_j\mbox{d}^3y\ \ 
\Psi^{\dagger}\left({\bf r}_1,\cdots,{\bf x}_i-\frac{{\bf y}}{4},
\cdots,{\bf x}_j+\frac{{\bf y}}{4},\cdots,{\bf r}_A\right) \,\nonumber \\
& & \tilde{v}_{\pi,Rel}\left(|{\bf x}_i-{\bf x}_j|,y\right)\ \,
\Psi\left({\bf r}_1,\cdots,{\bf x}_i+\frac{{\bf y}}{4},
\cdots,{\bf x}_j-\frac{{\bf y}}{4},\cdots,{\bf r}_A\right),
\ea
where $\tilde{v}_{\pi,Rel}\left(|{\bf x}_i-{\bf x}_j|,y\right)$ is previously
calculated in Eq.\ (\ref{vnonlocal}). The integration over the ${\bf r}_k$'s,
${\bf x}_i$, ${\bf x}_j$ and the solid angle of ${\bf y}$ is carried out
by the Monte Carlo method, while that over the magnitude of $y$ is carried
out with Gauss-Laguerre integral.

\subsection{VMC results}

The VMC results for $^3$H and $^4$He are listed in
Table IV. Note that in principle we should use GFMC to calculate the
exact binding energies, but the relativistic
effects resulting from the difference between $\langle H_R\rangle$ and
$\langle H_{NR}\rangle$ are small and presumably not too different from
those estimated using VMC.

The total relativistic effect on the binding energy is $\sim$0.3 MeV
for $^3$H and $\sim$1.8 MeV for $^4$He. Most of the effect comes from the
boost correction which is 0.42 MeV for $^3$H and 1.94 MeV for $^4$He.
The net effect of relativistic corrections to the kinetic energy and 
the two-body potential, on the binding energy is rather small:
$\sim -0.12\pm 0.06$ MeV in ($^3$H) and
$\sim -0.17\pm 0.10$ MeV in ($^4$He). 
Since both $H_{NR}$ and $H_R$ are constrained to give
the same deuteron binding energy, the changes in $\langle T\rangle$ and
$\langle\tilde{v}_{ij}\rangle$ cancel exactly in $^2$H.
In $^3$H and $^4$He they appear to largely cancel and 
give a rather small net effect.

In view of the slow convergence of the $\langle | \tilde{v}_{\pi,Rel} - 
\tilde{v}_{\pi,NR}| \rangle$, when expanded in powers of $p^2/m^2$, one 
may question the validity of calculating the boost interaction $\delta v$ 
only up to first order in $P^2/4m^2$.  The average  
kinetic energies of nucleons in nuclei are rather small giving 
average $p^2_i/m^2 < 0.1$.  The expansion has good convergence for 
such values. However, two
nucleons can occasionally have large relative momenta 
when they come close together as illustrated
in Fig.~\ref{fig:3ball}. In that configuration ${\bf p}_i \sim {\bf p}$ and
${\bf p}_j \sim -{\bf p}$ are both large due to
strong short-range interaction between nucleons $i$ and $j$.
Such configurations are responsible for the slow convergence of 
the expansion of
$\langle | \tilde{v}_{\pi,Rel} - \tilde{v}_{\pi,NR}| \rangle $.
In these configurations the squares of the total pair 
momenta are of order $P^2_{ik} \sim P^2_{jk} \sim p^2$, while $P^2_{ij}$ 
has small, near average value.  Thus the expansion parameter 
$P^2/4m^2$ for the boost interaction is effectively four times smaller than 
that of $(\tilde{v}_{\pi,Rel} - \tilde{v}_{\pi,NR})$ when the momenta 
are generated by pair correlations, therefore we 
expect the boost expansion to converge more rapidly.  Moreover, 
$\langle | \tilde{v}_{\pi,Rel} - \tilde{v}_{\pi,NR}| \rangle $
is much larger ($\sim$ 6 and 13 MeV in $^3$H and $^4$He respectively) 
than $\langle | \delta v | \rangle $
and has to be calculated with higher relative accuracy to obtain a 
final total energy with error $\sim$ 1 \%.

\section{Conclusions and Outlook}

We find that the relativistic effects in on-shell OPEP are quite substantial.
The expectation values of $\tilde{v}_{\pi,Rel}$ are smaller than those 
of $\tilde{v}_{\pi,NR}$ by $\sim$ 15 \%.  Since the expectation values 
of OPEP are much larger than nuclear binding energies, the differences 
in the OPEP expectation values are comparable to the total nuclear energy. 
However, nuclear Hamiltonians are not derived from first principles, 
they are obtained by fitting data.  The substantial difference between 
$\tilde{v}_{\pi,Rel}$ and $\tilde{v}_{\pi,NR}$ is compensated in the
$H_{NR}$ by that in the kinetic energy $T$ and $\tilde{v}_R$ so that it 
gives the same scattering cross sections and deuteron energy as the 
$H_R$.  We find that this compensation works rather well for three- and 
four-body nuclei.  In absence of the boost interaction our nonrelativistic
and relativistic Hamiltonians seem to give very similar results for the 
binding energies and wave functions of light nuclei. It is probably 
necessary to examine one-pion exchange current contributions to 
elastic scattering form factors and radiative capture reactions to
see the effect of the $m/E$ factors in the OPEP.

The modern two-nucleon potential models, which include the Nijmegen models 
I, II and Reid-93~\cite{SKTS94}, Argonne $v_{18}$~\cite{Wiringa95} and 
CD-Bonn~\cite{Machleidt96}, accurately reproduce the NN-scattering data 
in the Nijmegen data base.  
Friar {\it et al.}~\cite{Friar93} have studied the triton energy with
the Nijmegen and Argonne models, without boost or three-nucleon 
interactions, using accurate Faddeev calculations.  The energies 
obtained with the three local potential models, Reid-93, Nijmegen-II 
and Argonne $v_{18}$ are respectively -7.63, -7.62 and -7.61 MeV.
These energies are very close, and these models also give very similar 
values (5.70, 5.64 and 5.76 \%) 
for $P_D$, the fraction of D-state in the deuteron.
The boson exchange Nijmegen-I model contains nonlocal terms and gives 
-7.72 MeV for triton energy and 5.66\% for $P_D$.  Comparison of the results of 
Nijmegen I and II models indicates that the total effect of the 
nonlocalities on energies and the wave functions could be small.
The present results support this conclusion; inclusion of relativistic 
nonlocalities in on-shell OPEP and kinetic energy lowers the triton 
energy by $\sim$ 0.1 MeV and $P_D$ by 0.04\%.

In contrast the CD-Bonn potential gives rather different results from the 
Nijmegen and Argonne models.  It gives a triton energy of -8.00 MeV and 
$P_D$ = 4.83\%.  The OPEP in the CD-Bonn model has additional off-shell 
nonlocalities which disappear in the on-shell relativistic OPEP.  It is 
defined as~\cite{Machleidt}:
\ba
\tilde{v}_{\pi,CDB}({\bf p}^{\prime},{\bf p}) &=&
- \frac{f^2_{\pi NN}}{\mu^2}\ \frac{\boldtau_i \cdot \boldtau_j}{\mu^2+q^2}
\ \frac{m}{E}\ \frac{m}{E^{\prime}}\ 
\left[ \frac{}{}\boldsigma_i \cdot {\bf q}\ \boldsigma_j \cdot {\bf q} \right.
\nonumber \\
& & \left. +(E^{\prime}-E) \left( \frac 
{\boldsigma_i \cdot {\bf p}\ \boldsigma_j \cdot {\bf p}}{E+m} - \frac 
{\boldsigma_i \cdot {\bf p}^{\prime}\ \boldsigma_j \cdot {\bf p}^{\prime}}
{E^{\prime}+m} \right) \right],
\ea
where $E=\sqrt{m^2+p^2}$, $E^{\prime}=\sqrt{m^2+p^{\prime 2}}$.
The term proportional to $(E^{\prime} - E)$ 
does not contribute to the on-shell OPEP, and is absent from our 
$\tilde{v}_{\pi,Rel}$  given by Eq. (\ref{opep}).  The OPEP gives large
contributions 
by coupling states with small $p$ to large $p^{\prime}$.  Therefore we  
consider the case $p = 0$ for which ${\bf p}^{\prime} =-{\bf q}$, and 
expand in powers of $q^2/m^2$.  This gives:
\ba
\tilde{v}_{\pi,CDB}({\bf q},0) &=& \tilde{v}_{\pi,NR} \left(1- 
\frac{3 q^2}{4 m^2} + \cdots \right),     \\
\tilde{v}_{\pi,Rel}({\bf q},0) &=& \tilde{v}_{\pi,NR} \left(1- 
\frac{q^2}{2 m^2} + \cdots \right),
\ea
indicating presence of larger relativistic corrections 
in the $\tilde{v}_{\pi,CDB}$.

The second order contributions 
of the $\tilde{v}_{\pi,CDB}$ provide a sightly better approximation to 
the sum of the twelve time-ordered two-pion exchange diagrams~\cite{Smith76}
with only positive energy nucleons in the intermediate states, in relativistic 
field theories with pseudoscalar coupling.
The two-pion exchange diagrams with antinucleons are discarded 
on arguments based on chiral symmetry.  It is not obvious that this 
off-shell term must be retained, and not discarded.  It is necessary to 
find experimental tests for its existence, as well as for the suppression 
of OPEP by $m/E$ factors considered in this work.

The off-shell behavior of the OPEP can also be changed by using combinations 
of pseudoscalar and pseudovector couplings.  In Friar's notation~\cite{Friar80}
the possible off-shell behaviors are characterized with 
parameters $\tilde{\mu}$ and $\nu$, and up to order $p^2/m^2$ they are 
related by unitary transformations.  Up to this order our $\tilde{v}_{\pi,Rel}$ 
has an off-shell behavior with $\nu=1/2$ and $\tilde{\mu}=0$, while that of 
$\tilde{v}_{\pi,CDB}$ has $\nu=1/2$ and $\tilde{\mu}=-1$.  The $P_D$ has 
smaller values for $\tilde{\mu}=-1$~\cite{Adam93} used in the CD-Bonn potential.

The boost interaction $\delta v$ gives the dominant relativistic 
correction to the binding energies of light nuclei in the present 
formalism.  Contributions of $\delta v$ are very small, only $\sim$ 1 \% 
of that of $\tilde{v}$ in $^3$H and $^4$He.  However they are not 
canceled by relativistic effects in either $T$ or $\tilde{v}$, and 
therefore dominate the net effect.  Like the two-nucleon, the 
three-nucleon interaction is also not derived from first principles.  
Urbana models of $V_{ijk}$ contain two terms: the attractive two-pion 
exchange term $V^{2\pi}_{ijk}$ and a repulsive phenomenological 
term $V^R_{ijk}$.  Their strengths are chosen to 
reproduce the triton energy and the density of nuclear matter 
without considering any relativistic effects.
In light nuclei the repulsive $\delta v$ contribution is about 37 \% of
that of $V^R_{ijk}$.  Thus the strength of $V^R_{ijk}$ in $H_R$ has to 
be reduced by 37 \% to obtain the experimental energies of light nuclei. 
One could also choose to increase the strength of
$V^{2\pi}_{ijk}$ or some combination of increasing $V^{2\pi}_{ijk}$ and
decreasing $V^R_{ijk}$, but in this work we keep $V^{2\pi}_{ijk}$ unchanged.
The difference in the three nucleon interaction then compensates for 
the omission of $\delta v$ in conventional nonrelativistic nuclear 
Hamiltonians.  It appears that this compensation works rather well in 
light nuclei having up to eight nucleons~\cite{P98}, 
as well as in nuclear and neutron matter up to normal densities \cite{APR98}.  It presumably works also in heavier nuclei where it 
is not tested.  However, at several times nuclear matter densities, encountered 
in neutron stars, the effective three-nucleon interaction overestimates 
the $\delta v$ contribution significantly \cite{APR98}.

\begin{acknowledgements}
The authors would like to thank J. Carlson, R. Schiavilla for
many interesting discussions, R. B. Wiringa for
his help on the nonrelativistic variational wave function.
A.A. acknowledges the kind hospitality of the Physics Department of the
University of Illinois at Urbana-Champaign, where large part of this work has
been performed.  The calculations were performed on
IBM SP machines at Cornell Theory Center, and on Cray
supercomputers at Pittsburgh Supercomputing Center. This work was supported
by the U.S. National Science Foundation under Grant PHY94-21309, and the work
of A.A. by Universidade de Lisboa, Junta de Investiga\c c\~ao Cient\'{\i}fica e 
Tecnol\'{o}gica under contract No. PBIC/C/CEN/1108/92.
\end{acknowledgements}

\newpage
\begin{appendix}

\section{Relativistic OPEP in configuration space}

The third term in Eq.~(\ref{expansion}) gives the second-order relativistic
correction to OPEP and is denoted as $v^{(2)}$:
\beq
v^{(2)}({\bf x},{\bf y})=\int\frac{\mbox{d}^3 Q}
{(2\pi)^3}\frac{\mbox{d}^3 q}{(2\pi)^3}g(q,Q)\cos^2\theta_{qQ}
\tilde{v}_{\pi,NR}({\bf q})
\mbox{e}^{i({\bf Q}\cdot{\bf y}+{\bf q}\cdot{\bf x})}
\label{integral2}
\eeq
where $\theta_{qQ}$ is the angle between ${\bf q}$ and ${\bf Q}$, and
\beq
g(q,Q)=\frac{1}{2}\frac{m^2Q^2q^2}{(m^2+Q^2+\frac{q^2}{4})^3}
\eeq
Expressing $\cos^2\theta_{qQ}$ as
\beq
\cos^2\theta_{qQ} = \frac{2}{3}\mbox{P}_2(\cos{\theta_{qQ}})+\frac{1}{3},
\eeq
we get
\ba
v^{(2)}({\bf x},{\bf y})&=&\frac{1}{3}\int\frac{\mbox{d}^3 Q}
{(2\pi)^3}\frac{\mbox{d}^3 q}{(2\pi)^3}\,g(q,Q)\,\tilde{v}_{\pi,NR}({\bf q})
\,\mbox{e}^{i({\bf Q}\cdot{\bf y}+{\bf q}\cdot{\bf x})} \nonumber \\
&+&\frac{2}{3}\int\frac{\mbox{d}^3 Q}
{(2\pi)^3}\frac{\mbox{d}^3 q}{(2\pi)^3}\,g(q,Q)\,\mbox{P}_2(\cos{\theta_{qQ}})
\,\tilde{v}_{\pi,NR}({\bf q})\,\mbox{e}^{i({\bf Q}\cdot{\bf y}+{\bf q}
\cdot{\bf x})}.
\label{secondterm}
\ea
The first integral denoted by $v^{(2)}_1({\bf x},{\bf y})$
is independent of $\theta_{qQ}$ and can be easily
evaluated by using Eq.~(\ref{vpift}) and the following identities:
\ba
& &e^{i{\bf q}\cdot{\bf r}}=4\pi\sum_{lm}i^lY_{lm}^*(\hat{q})Y_{lm}(\hat{r})
j_l(qr),\label{math1} \\
& &\int Y_{lm}^*(\hat{q})Y_{l^{\prime}m^{\prime}}(\hat{q})\mbox{d}\Omega_q=
\delta_{ll^{\prime}}\delta_{mm^{\prime}},\label{math2} \\
& &\int v(r)\boldsigma_i\cdot\hat{r}\boldsigma_j\cdot\hat{r}\ e^{i{\bf q}
\cdot{\bf r}}\mbox{d}^3r=-\boldsigma_i\cdot\boldnabla_q\boldsigma_j\cdot
\boldnabla_q\int v(r)e^{i{\bf q}\cdot{\bf r}}\frac{1}{r^2}\mbox{d}^3r.
\label{math3}
\ea
We obtain:
\ba
v^{(2)}_1({\bf x},{\bf y})=\frac{\mu}{9\pi^3}\frac{f_{\pi NN}^2}{4\pi}
\left[\int Q^2\mbox{d}Q
q^2\mbox{d}q\,g(q,Q)\,{\cal Y}_{\pi}(q)\,j_0(Qy)j_0(qx)\,\boldsigma_i\cdot
\boldsigma_j \ \ \ \ \ \ \ \ \ \ \ \right. \nonumber \\
\hspace{1in} \left. +\int Q^2\mbox{d}Qq^2\mbox{d}q\,g(q,Q)\,{\cal T}_{\pi}(q)\,j_0(Qy)j_2(qx)
\,S_{ij}(\hat{x},\hat{x})\right]\boldtau_i\cdot\boldtau_j
\label{v21}
\ea
where ${\cal Y}_{\pi}(q)$ and ${\cal T}_{\pi}(q)$ are previously given in
Eqs.~(\ref{yft}) and (\ref{tft}).

To calculate the second integral in Eq.~(\ref{secondterm}), we use
\beq
\mbox{P}_l(\cos\theta_{qQ})=\frac{4\pi}{2l+1}\sum_mY_{lm}^*(\hat{Q})
Y_{lm}(\hat{q}).
\eeq
The integral over the solid angles becomes:
\beq
\int \mbox{d}\Omega_Q\mbox{d}\Omega_q e^{i({\bf Q}\cdot{\bf y}+{\bf q}\cdot
{\bf x})}\mbox{P}_2(\cos{\theta_{qQ}})=(4\pi)^2\mbox{P}_2(\cos{\theta_{xy}})
j_2(Qy)j_2(qx),
\eeq
and the second term in Eq. (\ref{secondterm}), denoted by
$v^{(2)}_2({\bf x},{\bf y})$ is:
\ba
v^{(2)}_2({\bf x},{\bf y})&=&\frac{2\mu}{9\pi^3}\frac{f_{\pi NN}^2}{4\pi}
\left\{\mbox{P}_2(\cos{\theta_{xy}})\!\int\!Q^2\mbox{d}Qq^2\mbox{d}qg(q,Q)
\left[{\cal Y}_{\pi}(q)+{\cal T}_{\pi}(q)\right]j_2(Qy)j_2(qx)
\boldsigma_i\!\cdot\!\boldsigma_j, \right. \nonumber \\
&+& \left. 3\boldsigma_i\!\cdot\!\boldnabla_x\boldsigma_j\!\cdot\!\boldnabla_x
\!\left[\mbox{P}_2(\cos{\theta_{xy}})\!\int\!Q^2\mbox{d}Q\mbox{d}qg(q,Q)
{\cal T}_{\pi}(q)j_2(Qy)j_2(qx)\right]\right\}\boldtau_i\!\cdot\!\boldtau_j.
\label{v22}
\ea
Here $\theta_{xy}$ is the angle between ${\bf x}$ and ${\bf y}$. 
The gradient operators $\boldnabla_x$ in the second term act on both
$\mbox{P}_2(\cos{\theta_{xy}})$ and $j_2(qx)$.

The $Q$-integral in Eqs.~(\ref{v21}) and (\ref{v22}) can be performed analytically and results
\ba
\int Q^2\mbox{d}Q\,g(q,Q)\,j_0(Qy) &=& f(y) \left[Z_2(q,y)-Z_1(q,y))\right], \\
\int Q^2\mbox{d}Q\,g(q,Q)\,j_2(Qy) &=& f(y) Z_1(q,y),
\ea
where f(y) is the Yukawa function given in Eq.(2.33) and
\ba
Z_1(q,y) &=& \frac{(\pi qy)^2}{8}\,\mbox{e}^{-(\sqrt{m^2+
\frac{q^2}{4}} - m)\,y}, \\
Z_2(q,y) &=& \frac{3}{\sqrt{m^2+\frac{q^2}{4}}\ y}\ Z_1(q,y).
\ea
$v^{(2)}$ is finally obtained as:
\ba
v^{(2)}({\bf x},{\bf y}) = \frac{\mu}{3}\frac{f_{\pi NN}^2}{4\pi}f(y)[
& & I_1(x,y,\theta_{xy})\,\boldsigma_i\cdot\boldsigma_j +
    I_2(x,y,\theta_{xy})\,S_{ij}(\hat{x},\hat{x}) + \nonumber \\ 
& & I_3(x,y,\theta_{xy})\,S_{ij}(\hat{y},\hat{y}) +
    I_4(x,y,\theta_{xy})\,S_{ij}(\hat{x},\hat{y})]\boldtau_i\cdot\boldtau_j,
\ea
where the tensor operator $S_{ij}(\hat{x},\hat{y})$ is defined as:
\beq
S_{ij}(\hat{x},\hat{y}) = \frac{3}{2}\left(\boldsigma_i\cdot\hat{x}
\boldsigma_j\cdot\hat{y}+\boldsigma_i\cdot\hat{y}\boldsigma_j\cdot\hat{x}
\right) - \hat{x}\cdot\hat{y}\,\boldsigma_i\cdot\boldsigma_j,
\eeq
and $I_1$, $I_2$, $I_3$ and $I_4$ are given by:

\ba
I_1(x,y,\theta_{xy}) &=& {\cal F}^{\cal Y}_{2\,0}(x,y) - 
                         {\cal F}^{\cal Y}_{1\,0}(x,y) +
                      2\,{\cal F}^{\cal Y}_{1\,2}(x,y)\,
                                                 \mbox{P}_2(\cos\theta_{xy})\\
I_2(x,y,\theta_{xy}) &=& {\cal F}^{\cal T}_{2\,2}(x,y) -   
                      3\,{\cal H}^{\cal T}_{1\,3}(x,y) +
                      3\,{\cal F}^{\cal T}_{1\,4}(x,y)\, \cos^2\theta_{xy} \\
I_3(x,y,\theta_{xy}) &=& 6\,{\cal H}^{\cal T}_{1\,2}(x,y)  \\
I_4(x,y,\theta_{xy}) &=& -12\,{\cal H}^{\cal T}_{1\,3}(x,y)\cos\theta_{xy}  
\ea

where

\ba
{\cal F}^{\cal Y}_{\alpha\,l}(x,y) &=& \frac {1}{3\pi^3} \int q^2\mbox{d}q   
                                 Z_{\alpha}(q,y)\,{\cal Y}_{\pi}(q)\,j_l(qx) \\
{\cal F}^{\cal T}_{\alpha\,l}(x,y) &=& \frac {1}{3\pi^3} \int q^2\mbox{d}q 
                                 Z_{\alpha}(q,y)\,{\cal T}_{\pi}(q)\,j_l(qx) \\
{\cal H}^{\cal T}_{\alpha\,l}(x,y) &=& \frac {1}{3\pi^3}  \int q^2\mbox{d}q 
                                 Z_{\alpha}(q,y)\,{\cal T}_{\pi}(q)\,
                                        \frac {j_l(qx)} {(qx)^{4-l}} 
\ea

\end{appendix}

\newpage

\begin{figure}
\caption{Phase shifts as a function of lab energy. $\diamond$: the reference
phase shifts obtained with $H_{NR}$;\ \ $+$: those using $\tilde{v}_R$ from
$H_{NR}$ in $H_R$;\ \ $\Box$: those with the new relativistic
Hamiltonian $H_R$ with re-adjusted $\tilde{v}_R$.}
\label{fig:phase}
\end{figure}

\begin{figure}
\caption{Comparison of relativistic (solid lines) and nonrelativistic
(dashed lines) potentials $v_1-v_{14}$ of the operator format
(refer to Eq. \ref{av18op}). Note that $v_4$ and $v_6$ contain contributions
from both $\tilde{v}_{\pi}$ and $\tilde{v}_R$, and only a local
nonrelativistic OPEP is used.}
\label{fig:v14}
\end{figure}

\begin{figure}
\caption{Deuteron wave functions.}
\label{fig:wavefunc}
\end{figure}

\begin{figure}
\caption{Deuteron wave functions in momentum space.}
\label{fig:momdist}
\end{figure}

\begin{figure}
\caption{Ratio of deuteron relativistic to nonrelativistic
D-wave function in momentum space (solid line),
and the simple estimate of Eq.~\ref{ratio} (dashed line). The dotted and
dot-dashed lines represent results calculated using $T_{NR}$,
$\tilde{v}_{\pi,Rel}$ and $T_{Rel}$, $\tilde{v}_{\pi,NR}$ in $\phi_{D,Rel}$
in Eq. (\ref{ratio}), respectively. The wiggle in the calculated ratio
around 7 fm$^{-1}$ comes from a node in the wave function.}
\label{fig:ratio}
\end{figure}

\begin{figure}
\caption{$F_{\sigma\tau}$ and $F_{t\tau}$ as a function of x for various
values of y.}
\label{fig:f}
\end{figure}

\begin{figure}
\caption{VMC results for deuteron with 100,000 configurations.}
\label{fig:deutvmc}
\end{figure}

\begin{figure}
\caption{$\xi=\frac{f_R-f_{NR}}{f_{NR}}$ as a function of r.}
\label{fig:diffall}
\end{figure}

\begin{figure}
\caption{VMC results for triton with 50,000 configurations.}
\label{fig:tritonvmc}
\end{figure}

\begin{figure}
\caption{A diagram to illustrate the calculation of nonlocal interaction
contributions.}
\label{fig:nonlocal}
\end{figure}

\begin{figure}
\caption{A naive picture of $^3$H to illustrate large momentum contribution
from a configuration where two nucleons are close together.}
\label{fig:3ball}
\end{figure}

\newpage

\begin{table}
\caption{Nonrelativistic Green's Function Monte Carlo (GFMC) results (in MeV)
for light nuclei with the Argonne $v_{18}$ and Urbana IX potentials. The first
line gives the experimental energy while the next four list the calculated
total, kinetic, two- and three-body interaction energies. The last two lines
give the contribution of the pion exchange parts of 2- and 3-body interactions.}
\vspace{0.5cm}
\begin{tabular}{crrrrrr}
           & $^2$H  & $^3$H   & $^4$He   & $^6$Li   & $^7$Li   & $^8$Be\\ \hline
$E_{exp}$  &-2.2246   & -8.48   & -28.30   &-31.99    &-39.24    &-56.50     \\
$<E>$      &-2.2248(5)& -8.47(1)& -28.30(2)&-31.25(11)&-37.44(28)&-54.66(64) \\
$<T>$      &19.81     & 50.8(5) & 111.9(6) &150.8(10) &186.4(28) &246.3(56)  \\
$<v_{ij}>$ &-22.04    &-58.4(5) &-135.4(6) &-179.2(10)&-220.8(30)&-295.8(62) \\
$<V_{ijk}>$&   0.0    &-1.20(2) &  -6.4(1) &-7.2(1)   &-8.9(2)   &-14.8(5)   \\
$<v_{\pi}>$&-21.28    &-43.8(2) &-99.4(2)  &-128.9(5) &-152.5(7) &-224.1(9)  \\
$<V^{2\pi}>$& 0.0     & -2.17(1)& -11.7(1) &-13.5(1)  &-17.1(4)  &-28.1(8)   \\
\end{tabular}
\end{table}

\vspace{1cm}

\begin{table}
\caption{Deuteron properties}
\vspace{0.5cm}
\begin{tabular}{lrr}
                                & $H_{NR}$          &  $H_R$      \\ \hline
binding energy (MeV)$^\dagger$  & -2.242            & -2.242      \\
quadrupole moment (fm$^2$)      &  0.269            &  0.271      \\
\% of D state $(P_D)$           &  5.776            &  5.732
\end{tabular}
$\dagger$ without electromagnetic interactions
\end{table}

\vspace{1cm}

\begin{table}
\caption{Results of momentum space deuteron calculations.}
\vspace{0.5cm}
\begin{tabular}{crr}
         &$\langle\Psi_{NR}|H_{NR}|\Psi_{NR}\rangle$&$\langle\Psi_R|H_R|\Psi_R\rangle$   \\ \hline
$\langle E\rangle$  &  -2.242  \ \ \ \ \ \  &  -2.242 \ \ \ \    \\
$\langle T\rangle$  &  19.882  \ \ \ \ \ \  &  18.877 \ \ \ \    \\
$\langle\tilde{v}_{ij}\rangle$ & -22.125  \ \ \ \ \ \  & -21.119 \ \ \ \    \\
$\langle\tilde{v}_{\pi}\rangle$& -21.356  \ \ \ \ \ \  & -18.797 \ \ \ \    \\
$\langle\tilde{v}_R\rangle=\langle\tilde{v}_{ij}-\tilde{v}_{\pi}\rangle$& -0.769 \ \ \ \ \ \     &  -2.322 \ \ \ \    \\
$\langle\tilde{v}_{\pi,Rel}-\tilde{v}_{\pi,NR}\rangle$&               &   2.589 \ \ \ \   
\end{tabular}
\end{table}

\begin{table}
\caption{VMC results for $^3$H and $^4$He, calculated with 50,000
configurations.}
\vspace{0.5cm}
\begin{tabular}{crrrr|rrr}
           &        &  $^3$H$\ \ $ &      & &       & $^4$He$\ \ \ $  & \\
           &$H_{NR}$&  $H_R\ \ $  & $H_R-H_{NR}$& &$H_{NR}\ \ $& $H_R\ \ \ $   &$H_R-H_{NR}$\\ \hline
$\langle E\rangle^\dagger$&-8.24(3)&-7.94(4) &   0.30(5) & &-28.09(7)&-26.32(8)&1.8(1)\\
$\langle T\rangle$& 50.1(5)&48.6(5)  &   -1.5(7) & &104.8(9)&98.4(8) &-6(1)  \\
$\langle\tilde{v}_{ij}\rangle$ &-57.3(5)&-56.0(5) &  1.3(7) & &-127.6(9)&-121.5(9)& 6(1) \\
$\langle\tilde{V}_{ijk}\rangle$&-1.06(3)&-1.03(3) &     & &-5.29(9)&-5.20(8) &   \\
$\langle\delta v_{ij}\rangle$& &0.42(1)  &   & &        &1.94(3)  &   \\[2ex]
$\langle\tilde{V}_{ijk}^R\rangle$&0.98(3)&1.01(3)  &    & &5.26(7) &5.38(8)  &   \\
$\langle\tilde{v}_{\pi}\rangle$&-44.0(2)&-38.3(2) &   5.7(4)  & &-97.1(5)&-83.8(4) &13.3(1)
\end{tabular}
$\dagger$ without electromagnetic interaction
\end{table}

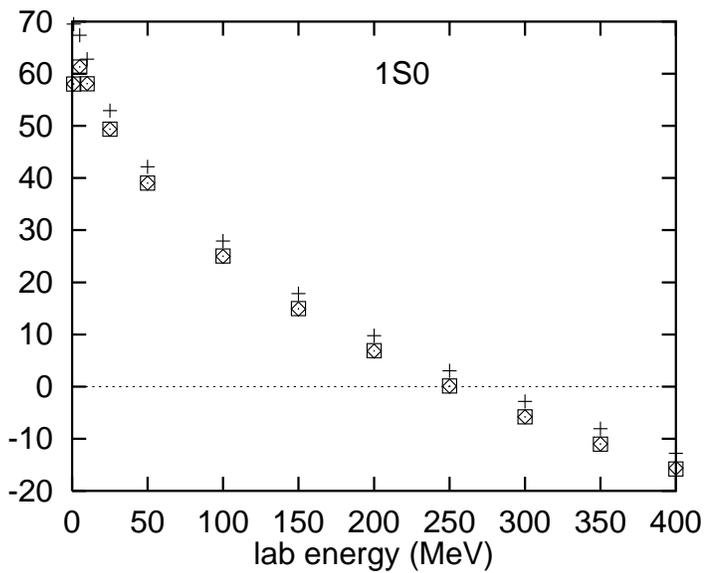
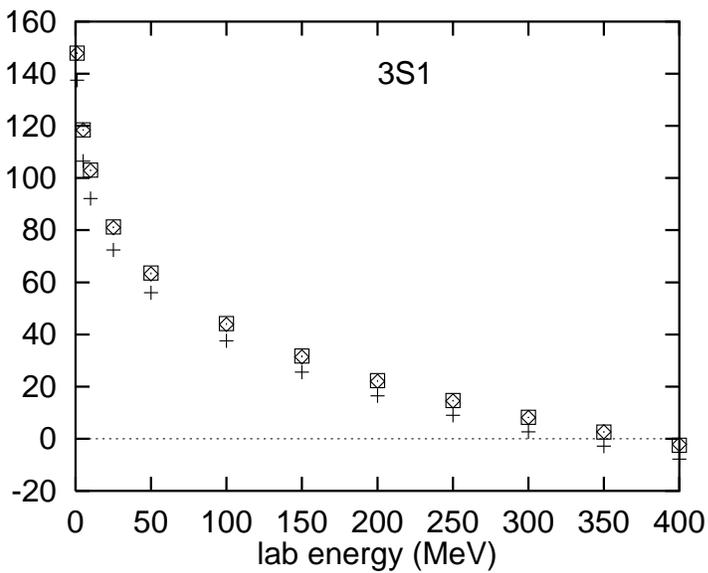
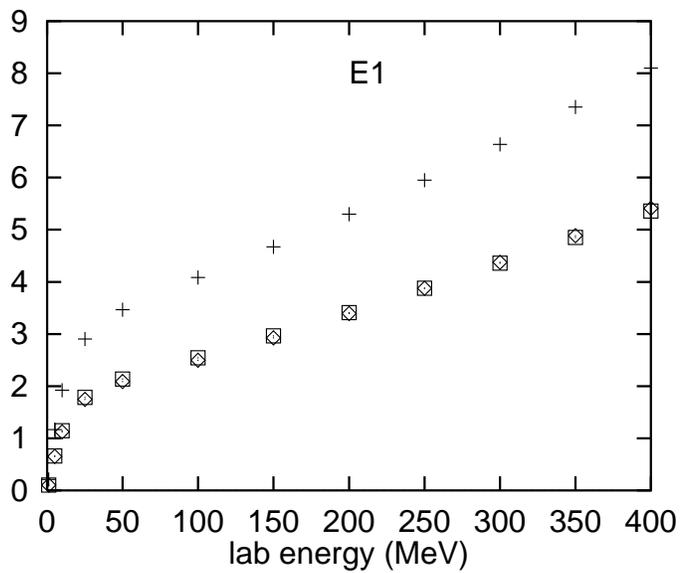
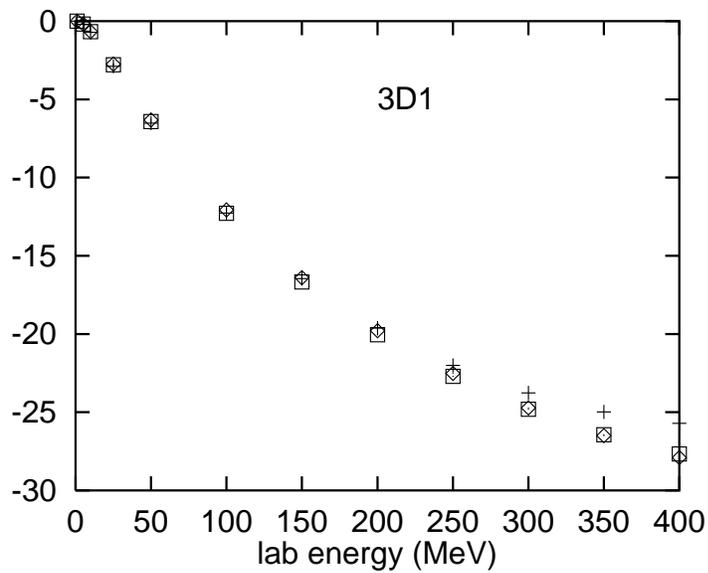
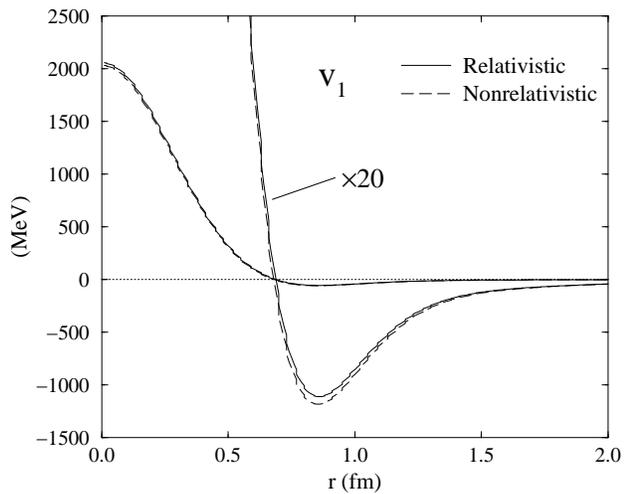
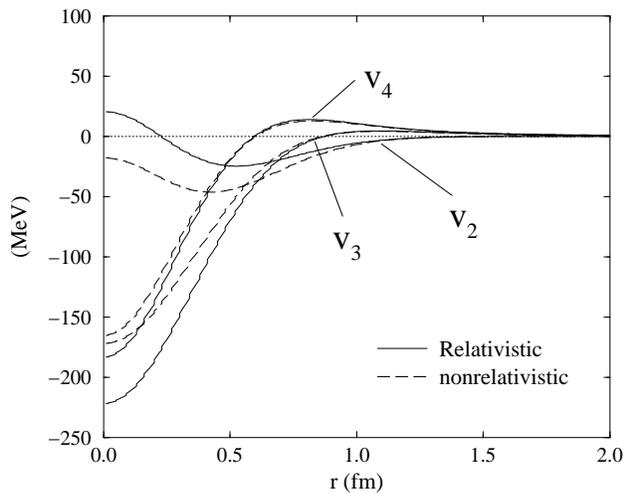
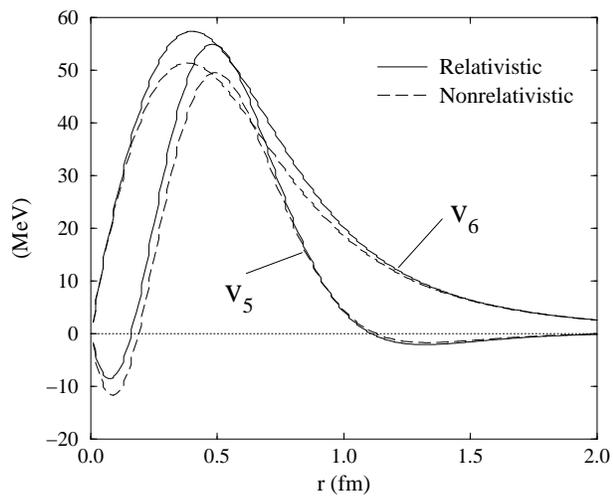
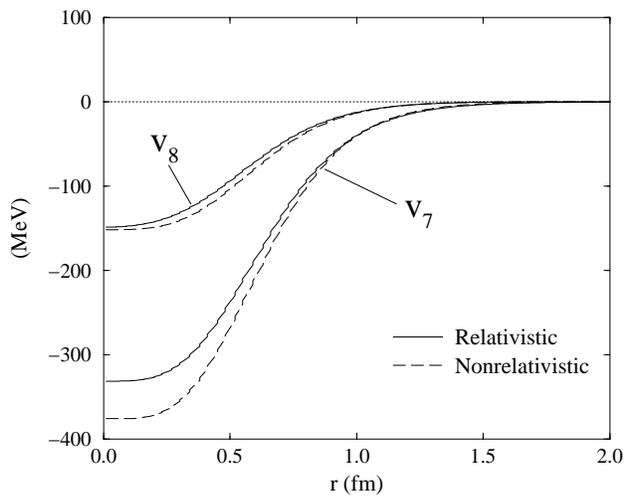
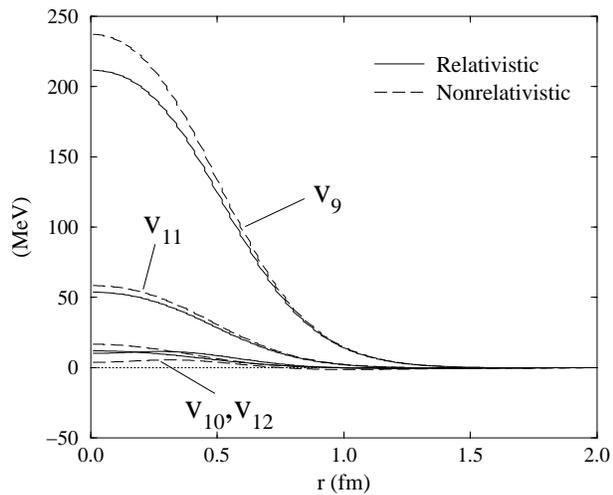
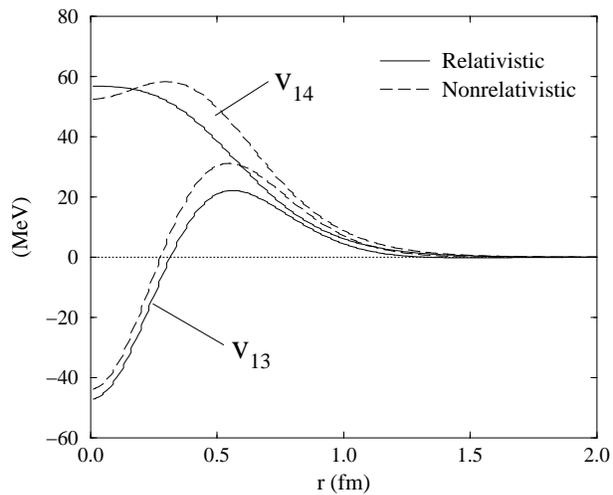
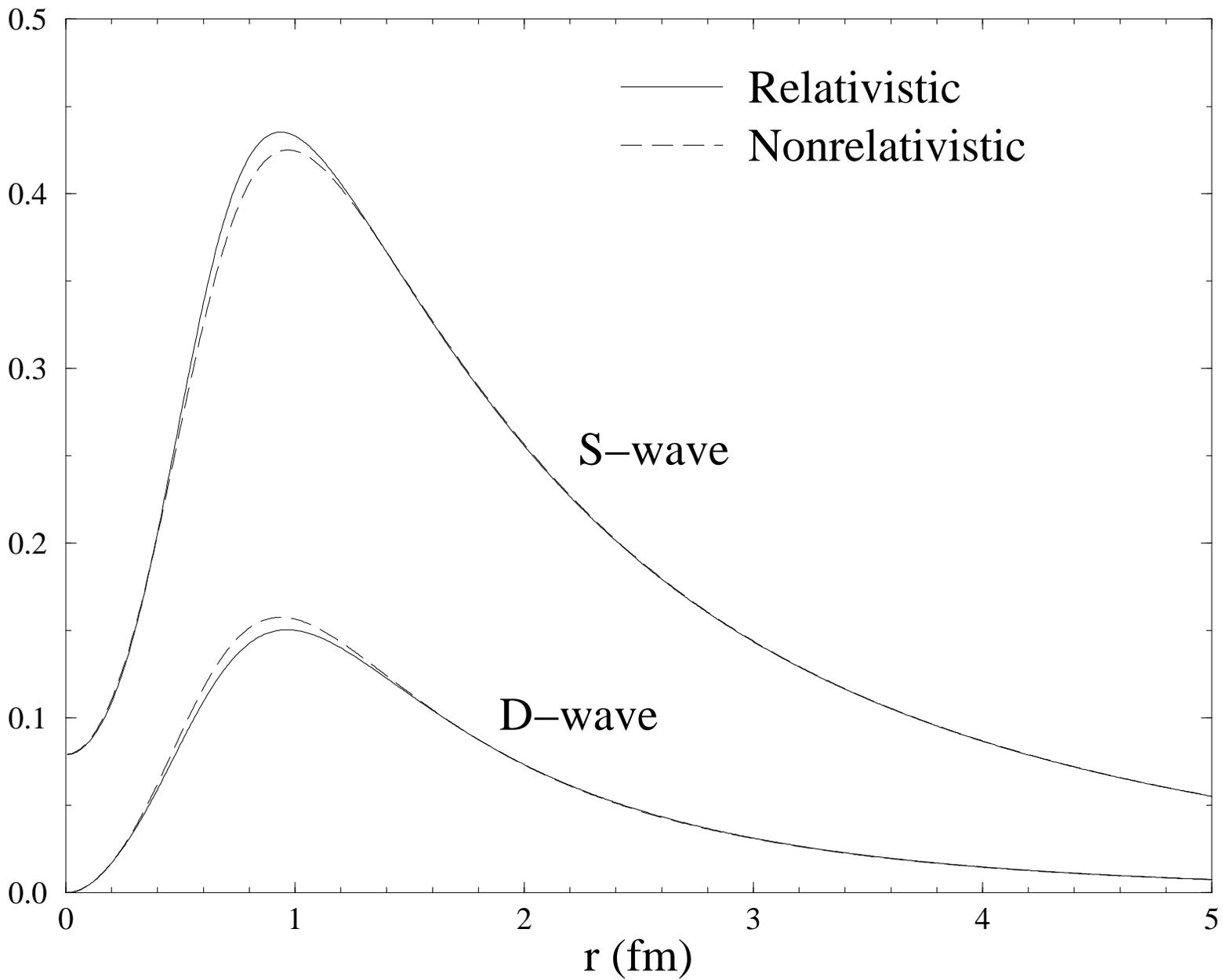
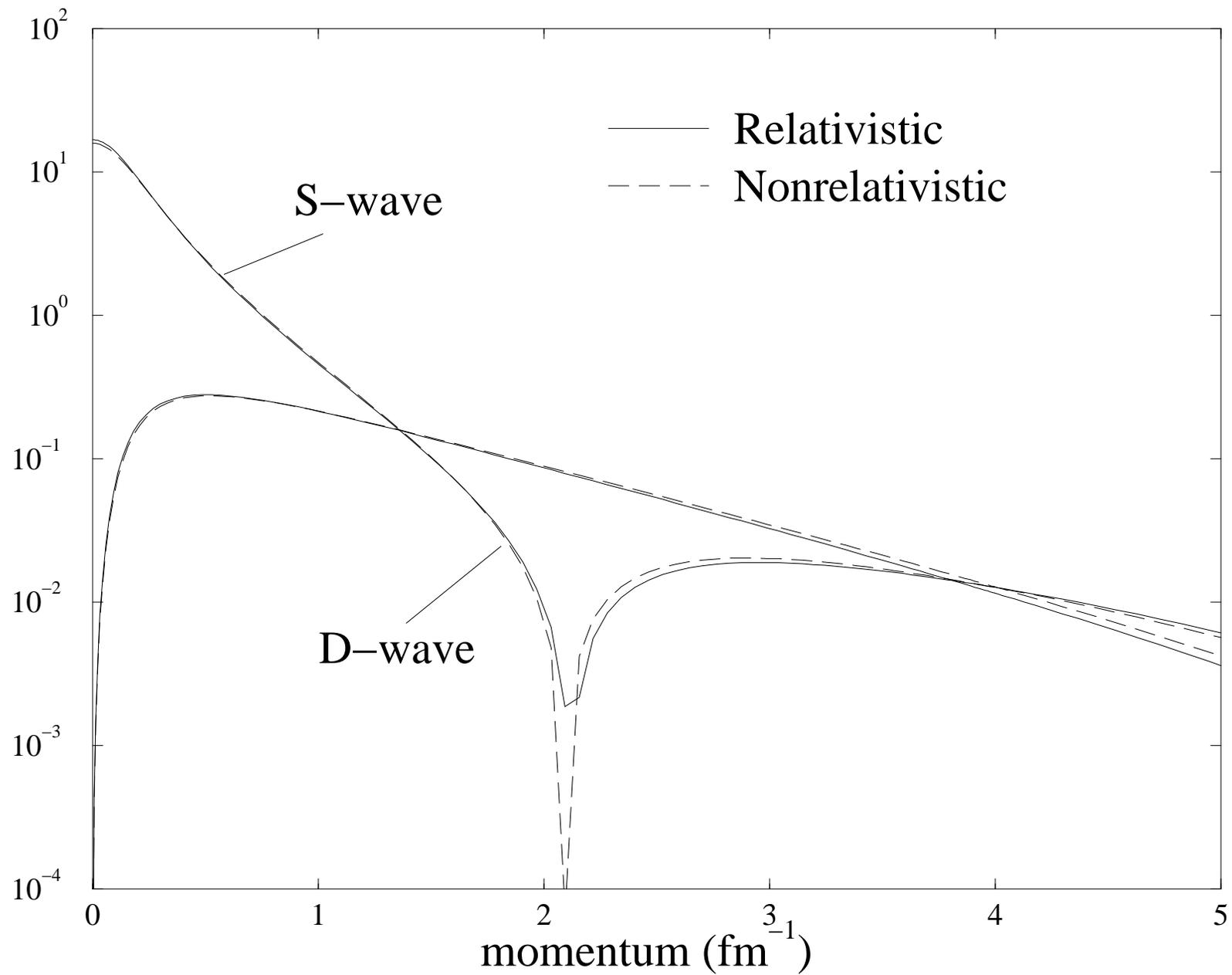
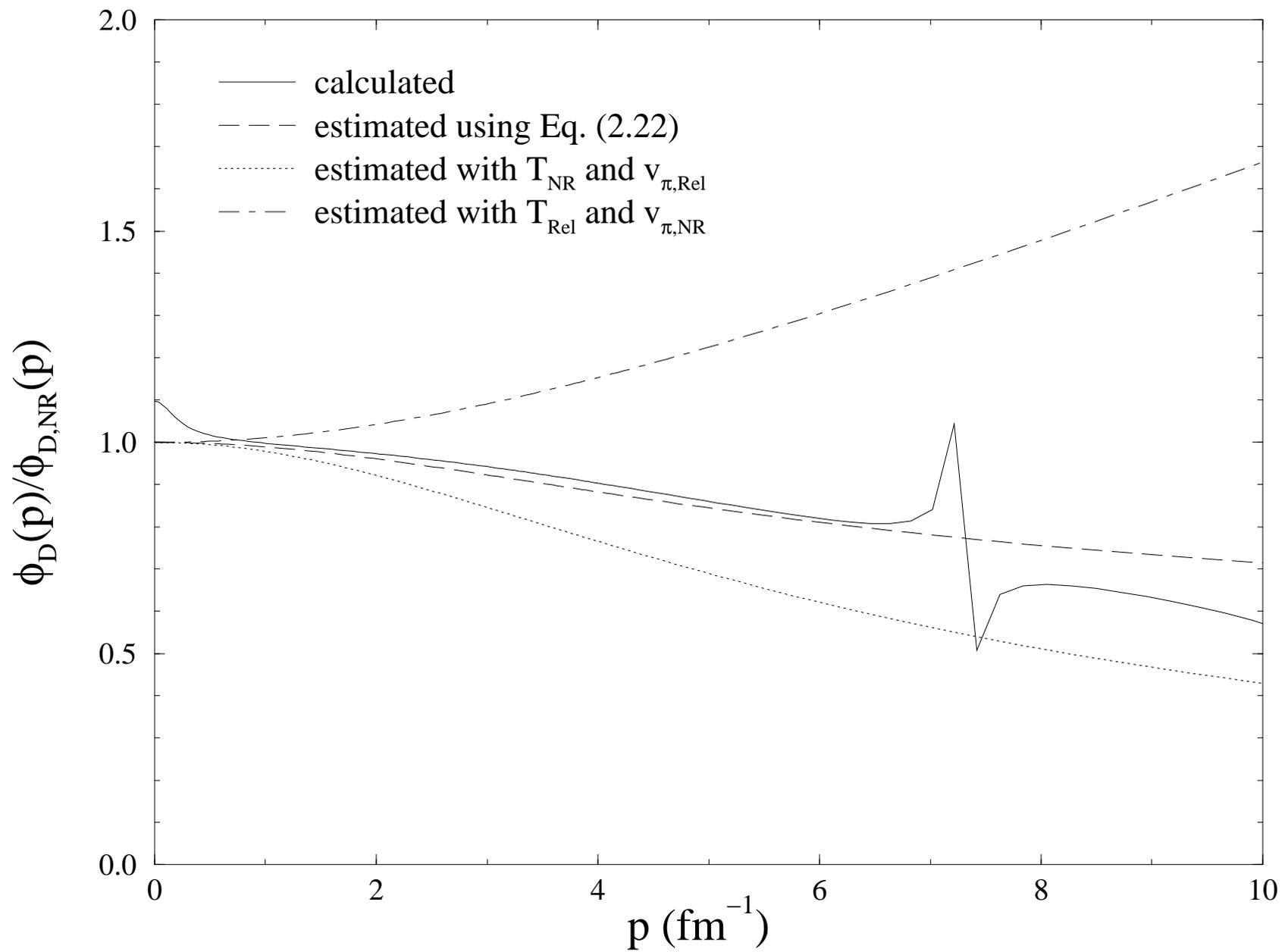
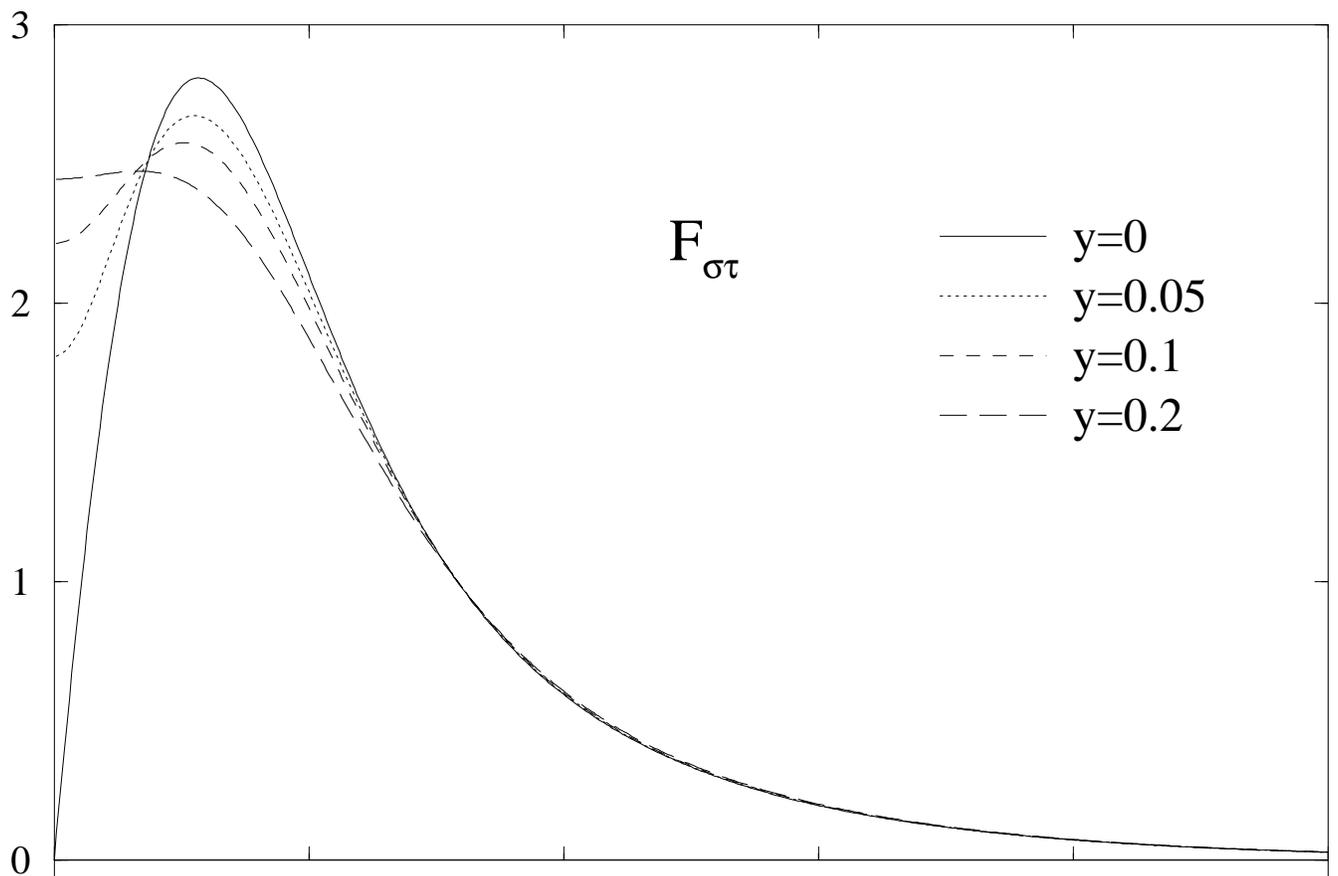
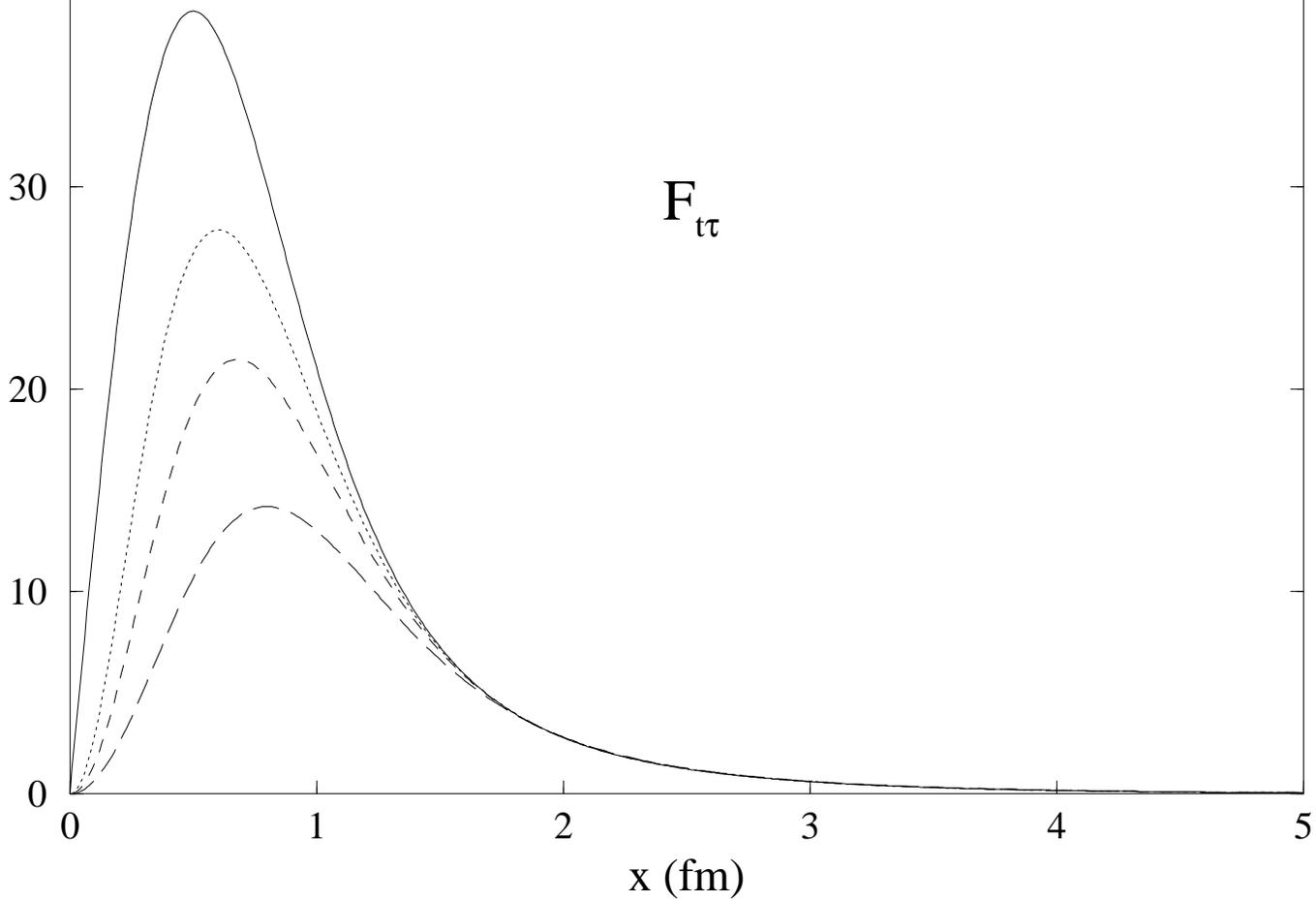
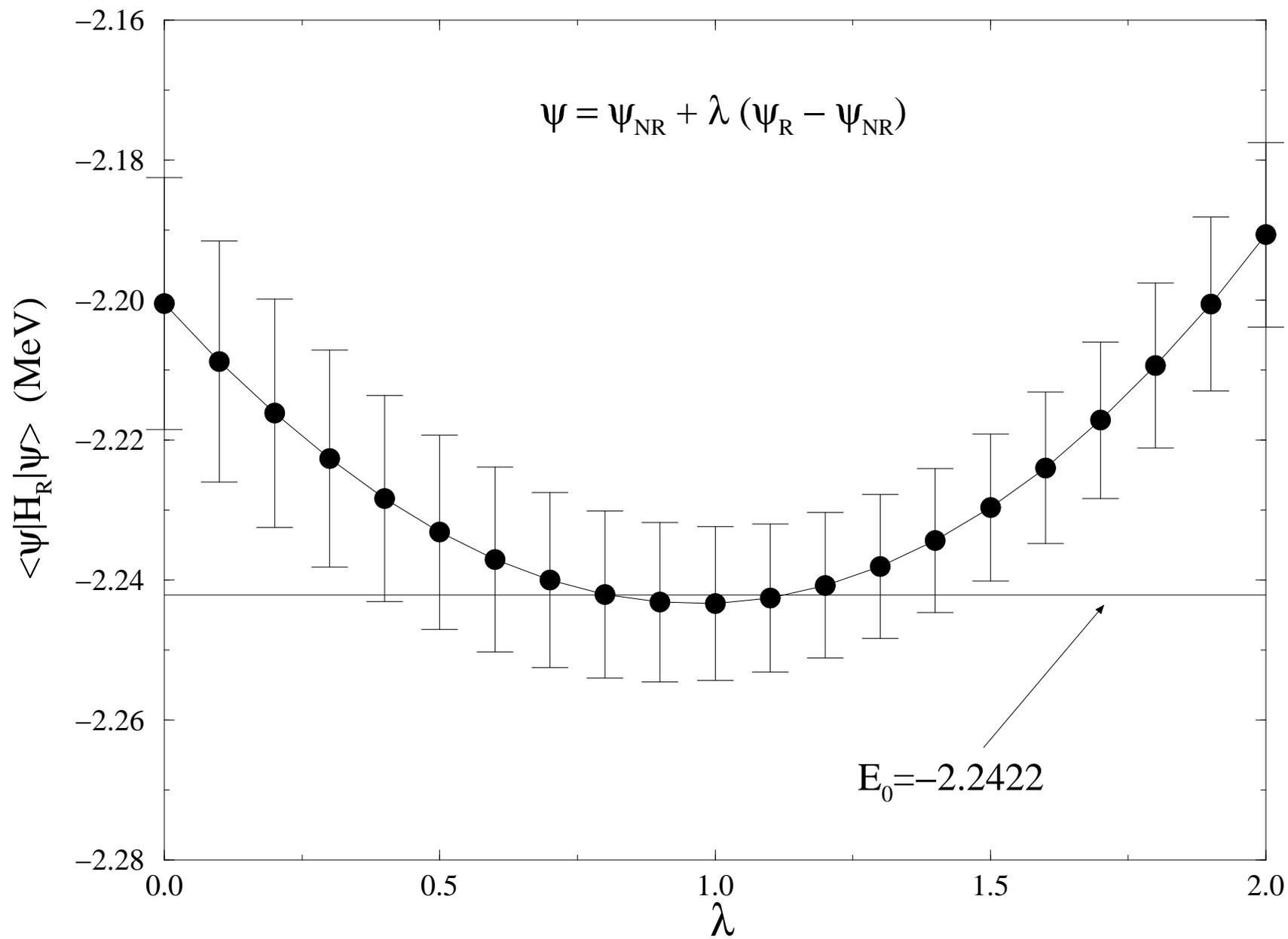
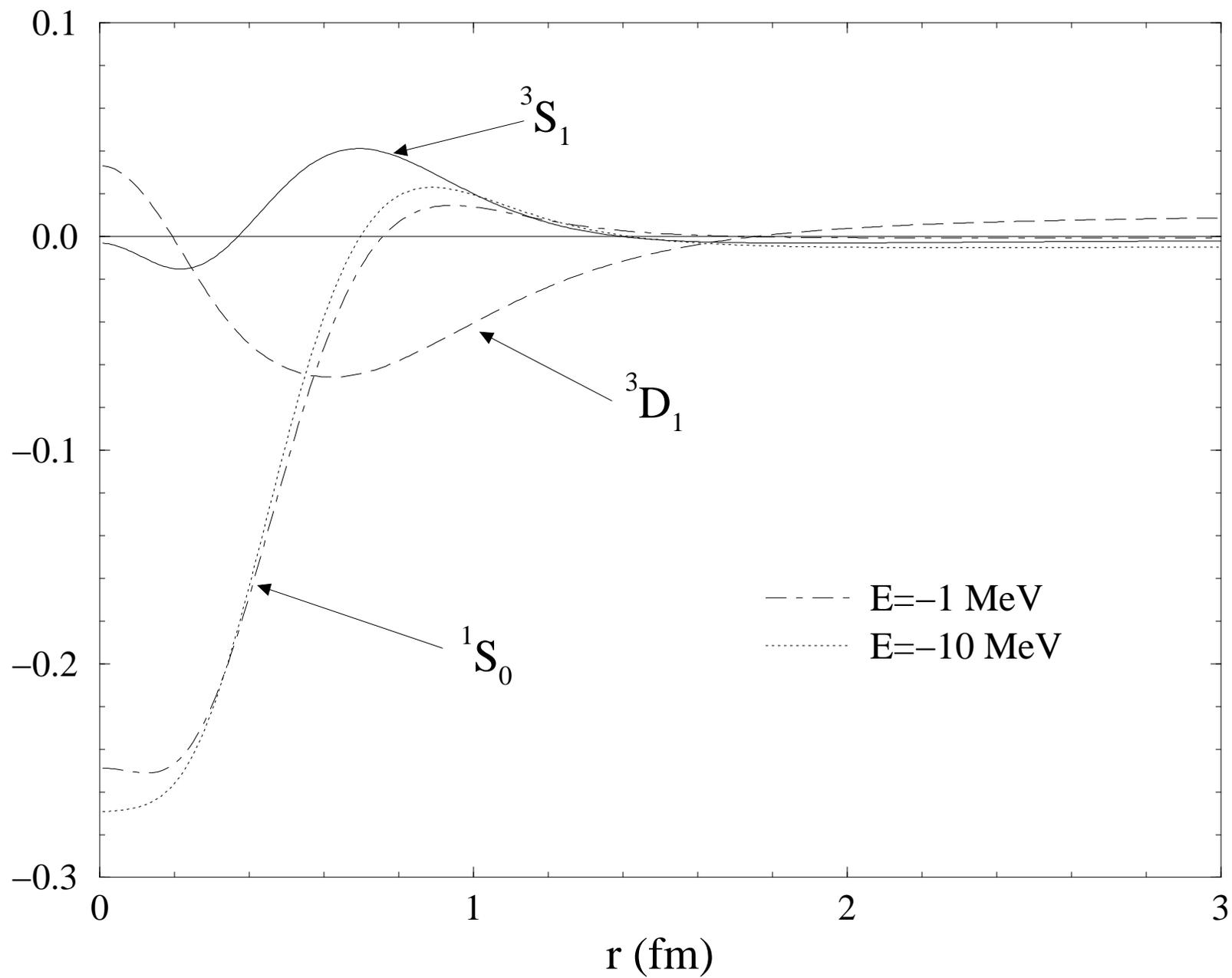
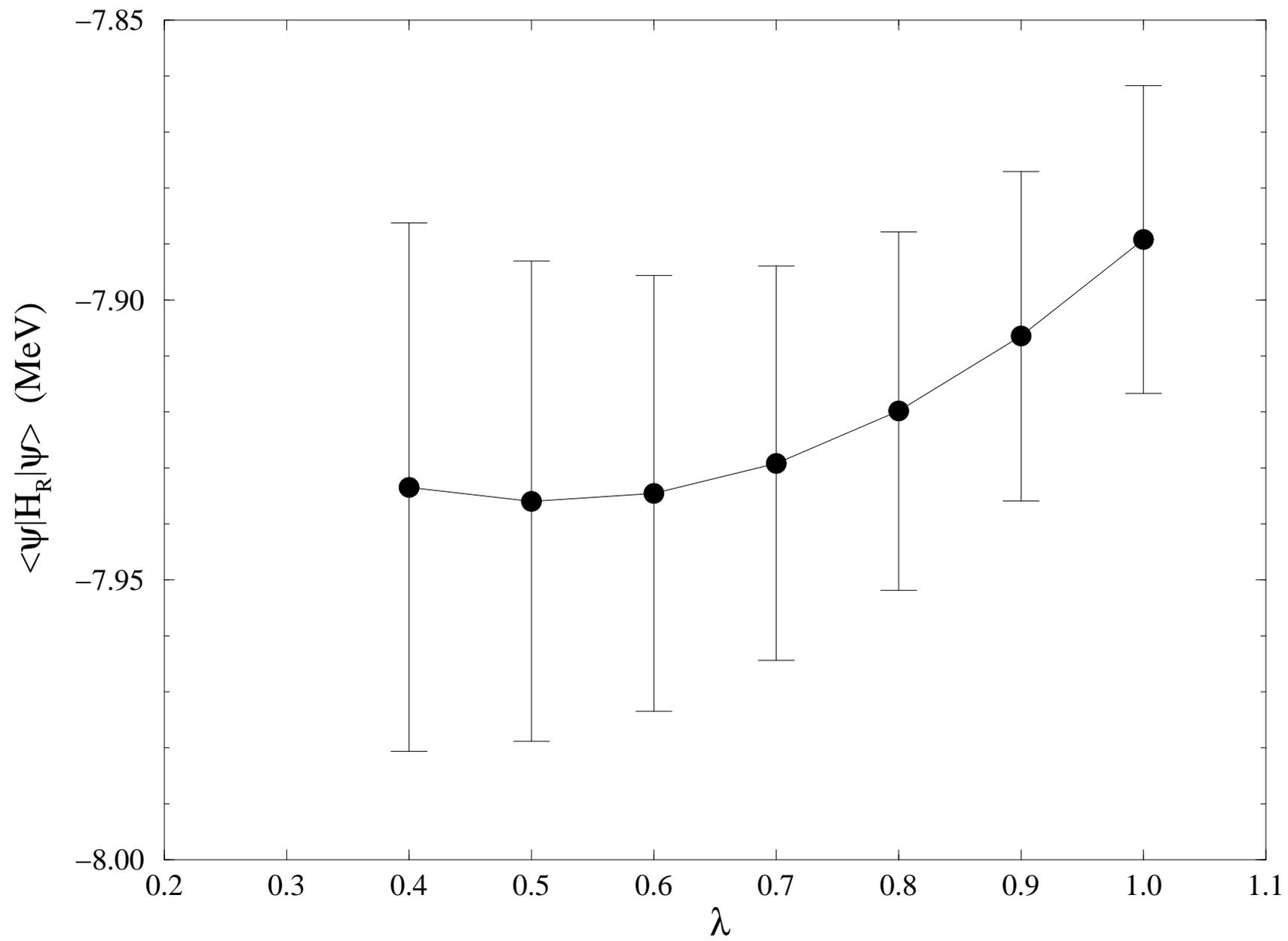
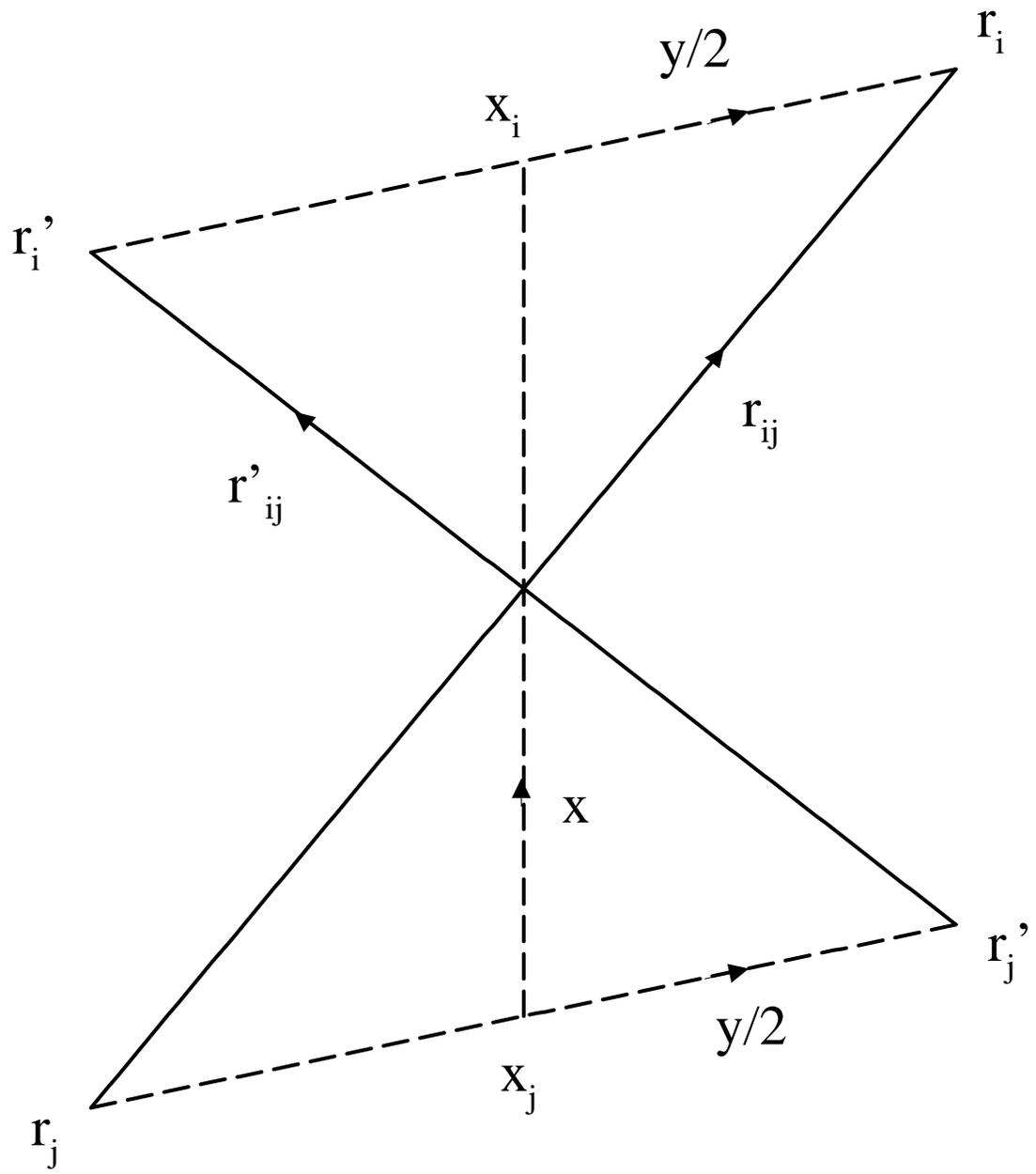
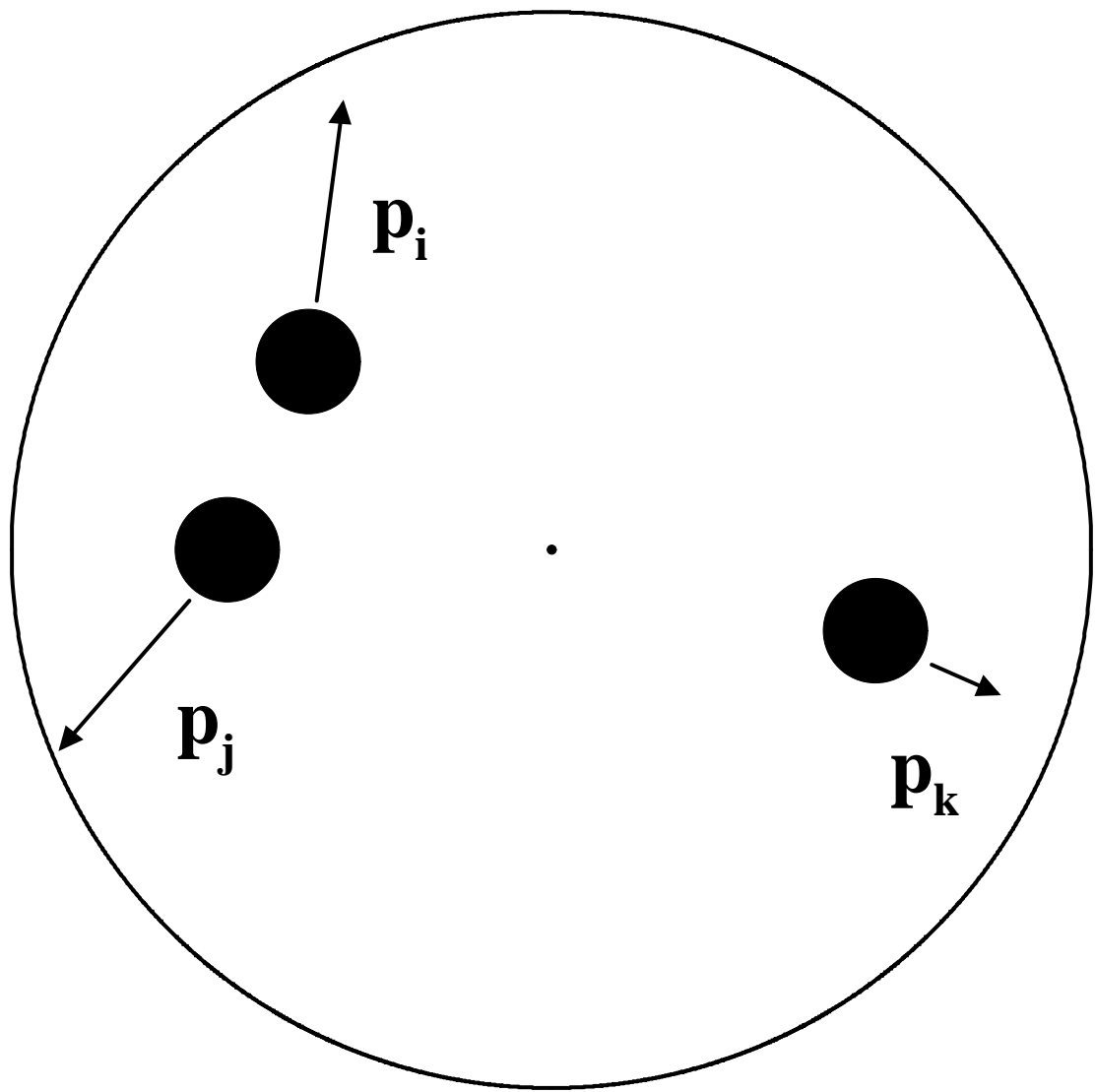
\end{document}